\documentclass[twocolumn,showpacs,preprintnumbers,amsmath,amssymb]{revtex4}
\usepackage{epsfig}
\usepackage{dcolumn}

\usepackage{graphicx}
\usepackage{dcolumn}
\usepackage{bm}

\newcommand{\be}{\begin{equation}}
\newcommand{\ee}{\end{equation}}
\newcommand{\bea}{\begin{eqnarray}}
\newcommand{\eea}{\end{eqnarray}}

\newcommand{\integer}{\relax{\rm I\kern-.18em N}}

\begin{document}

\title {Underlying gauge symmetries of second-class constraints systems}

\author{M. I. Krivoruchenko$^{1,2}$, Amand Faessler$^{2}$, A. A. Raduta$^{2,3,4}$, C. Fuchs$^{2}$}

\affiliation{
$^{1}$Institute for Theoretical and Experimental Physics$\mathrm{,}$
B. Cheremushkinskaya 25$\mathrm{,}$ 117259 Moscow, Russia \\
$^{2}$Institut f\"{u}r Theoretische Physik$\mathrm{,}$ Universit\"{a}t T\"{u}bingen$\mathrm{,}$
Auf der Morgenstelle 14$\mathrm{,}$ D-72076 T\"{u}bingen$\mathrm{,}$ Germany \\
$^{3}$Department of Theoretical Physics and Mathematics$\mathrm{,}$ Bucharest University$\mathrm{,}$ 
Bucharest$\mathrm{,}$ POBox MG11$\mathrm{,}$ Romania \\
$^{4}$Institute of Physics and Nuclear Engineering$\mathrm{,}$ Bucharest$\mathrm{,}$ POBox MG6$\mathrm{,}$ Romania \\
}

\begin{abstract}
Gauge-invariant systems in unconstrained 
configuration and phase spaces, equivalent to second-class 
constraints systems upon a gauge-fixing, are discussed. A mathematical pendulum on an 
$n-1$-dimensional sphere $S^{n-1}$ as an example of a mechanical second-class 
constraints system and the $O(n)$ non-linear sigma model as an example of a 
field theory under second-class constraints are discussed in details
and quantized using the existence of underlying dilatation gauge symmetry and by 
solving the constraint equations explicitly. 
The underlying gauge symmetries involve, in general, velocity dependent gauge 
transformations and new auxiliary variables in extended configuration space.
Systems under second-class holonomic constraints have gauge-invariant counterparts
within original configuration and phase spaces. The Dirac's supplementary conditions for wave functions 
of first-class constraints systems are formulated in terms 
of the Wigner functions which admit, as we show, a broad set of physically 
equivalent supplementary conditions. Their concrete form 
depends on the manner the Wigner functions are extrapolated from the constraint
submanifolds into the whole phase space.
\end{abstract}

\maketitle

\section{Introduction} 
\setcounter{equation}{0}

The specific feature of gauge theories is the occurrence of constraints
which restrict the phase space of the system to a submanifold. A systematic 
study of the Hamiltonian formulation of gauge
theories was made by Dirac \cite{DIRAC1,DIRAC2} who classified the
constraints and developed the operator quantization schemes of the
constraint Hamiltonian systems.

The Dirac's theory of constraint systems was combined further
with the Feynman path integral method. Faddeev \cite{FADD} found an explicit
measure on the constraint submanifolds, entering the path integrals. The Feynman diagram
technique for non-abelian gauge theories was
developed by Faddeev and Popov \cite{FAPO}. The effective QCD Lagrangian
involving ghost fields obeys the Becchi-Rouet-Stora (BRS) symmetry \cite{BRS}. 
A completely general
approach to quantization of gauge theories, in which the BRS transformation
acquires an intrinsic meaning, is developed by Fradkin and his collaborators 
\cite{FRVI,FRAD}. 
An alternative symplectic scheme has been proposed by Faddeev and Jackiw \cite{FAJA,JACK}. 
An introduction to quantization of the gauge theories with
the use of the path integral method can be found in Refs.\cite{HENN,SLFA}.
The
path integral representation for the evolution operator satisfies the unitarity
condition and meets requirements of the relativistic covariance. 

According to the Dirac's classification, constraint equations like $\Omega _{A}=0$,
appearing in the gauge theories, are of the first class. The Poisson
bracket of first-class constraint functions is a linear combination of constraint functions
\begin{equation}
\{\Omega _{A},\Omega _{B}\}=\mathcal{C}_{AB}^{\;\;\;\;D}\Omega _{D}
\label{FICLCO}
\end{equation}
where $\mathcal{C}_{AB}^{\;\;\;\;D}$ are some structure functions. The
quantization of gauge theories involves a set of gauge-fixing condition $%
\chi _{A}=0$. The functions $\chi _{A}$ must be admissible, i.e., the Poisson
bracket of the gauge-fixing and constraint functions is non-degenerate: 
\begin{equation}
\det \{\chi _{A},\Omega _{B}\}\ne 0.  \label{NONDEG}
\end{equation}
The gauge-fixing functions are independent
\begin{equation}
\{\chi _{A},\chi _{B}\}=0.  \label{COMPAR}
\end{equation}
Physical observables are gauge invariant and do not depend on the choice of
$\chi _{A}$. 

The wave functions satisfy the operator Dirac's supplementary
condition 
\begin{equation}
\hat{\Omega}_{B}\Psi =0.  \label{DSUPC}
\end{equation}

Constraint functions $\mathcal{G}_{a}$ of second-class constraints
systems obey the Poisson bracket relations which form a non-degenerate matrix 
\begin{equation}
\det \{\mathcal{G}_{a},\mathcal{G}_{b}\}\ne 0.  \label{FICL}
\end{equation}
The number of the constraints is even ($a=1,...,2m$).

Among physical systems which belong to the second-class constraints systems are 
anomalous gauge theories \cite{ANOM,AFAD,FSHA1,FSHA2,JOOO}, the $O(4)$ non-linear 
sigma model, which constitutes the lowest order of the chiral perturbation theory \cite{GALE}, 
many-body systems involving collective and independent-particle degrees of 
freedom \cite{MBODY}, and others.

The Dirac's quantization method of such systems consists in constructing
operators reproducing the Dirac bracket for canonical variables and taking
constraints to be operator equations of the corresponding quantum theory.

Batalin and Fradkin \cite{BATA} developed a quantization procedure for the 
second-class constraints systems, which converts constraints to the first class by
introducing new canonical variables. The problem reduces thereby to the
quantization of a first-class constraints system in an enlarged phase space.
This method (see also \cite{BATY}) was found to be particularly useful for construction of
the effective covariant Lagrangians in an extended configuration space \cite{BANE94,HONG04}.

The Hamilton-Jacobi scheme is also used for the quantization of constraint
systems \cite{GUEL,HAJA,PIME,BALE,HONG04}.

Gauge systems are quantized by imposing gauge-fixing conditions. The initial
system is reduced to a second-class constraints system. The
evolution operator in the path
representation can be written as \cite{FADD,AFAD} 
\begin{eqnarray}
Z &=&\int \ \prod_{t}\frac{d^{n}qd^{n}p}{(2\pi\hbar)^n}(2\pi\hbar)^{m}\prod_{a}\delta ({{\mathcal{G}}}_{a}){\sqrt{%
\det \{{\mathcal{G}}_{a},{\mathcal{G}}_{b}\}}} \nonumber \\
&\;& \times \exp \left\{ \frac{i}{\hbar}\int dt(p^{i}%
\dot{q}_{i}-{\mathcal{H})}\right\}  \label{PARQ2}
\end{eqnarray}
where ${\mathcal H}$ is the Hamiltonian.
The path integral representation (\ref{PARQ2}) allows not to solve the
constraint equations explicitly and work in the unconstrained phase space. 

The usefulness of the reduction of second-class constraints systems 
to equivalent gauged systems consists in getting 
the supplementary condition (\ref{DSUPC}) which is not evident otherwise. 

The second-class constraints systems to which gauge systems are reduced obey
specific requirements: 

(i) The constraint functions ${\mathcal{G}}_{a}$ split naturally into
canonically conjugate pairs ${\mathcal{G}_{a}=}(\chi _{A},\Omega _{B})$, 

(ii) the wave functions satisfy Eq.(\ref{DSUPC}), 

(iii) the gauge-fixing functions 
$\chi _{A}$ are identically in involution Eq.(\ref{NONDEG}), and

(iv) the constraint functions $\Omega _{B}$ associated to the gauge invariance 
are first class Eq.(\ref{FICLCO}).

As a consequence of Eq.(\ref{FICL}), the matrix $\{ \chi _{A},\Omega _{B} \}$ is non-degenerate.
The constraint functions are not defined uniquely. In particular, a linear transformations
of ${\mathcal{G}}_{a}$ by a non-degenerate matrix leads to constraint functions 
${\mathcal{G}}_{a}^{\prime}$ describing the same constraint submanifold.

{\it A question arises if constraints of an arbitrary second-class constraints 
system can be redefined to fulfill (i) - (iv)?}

This is the case for holonomic systems \cite{MFRF}.  
Such systems can be treated as being obtained upon 
a gauge-fixing of the corresponding gauge invariant systems.
Within the generalized Hamiltonian framework, constraints of holonomic 
systems are of the second class. In the Lagrangian formalism, the corresponding constraints
do not depend on generalized velocities. 

In this work, we review the underlying gauge symmetries of the holonomic 
systems and report new results connected to the quantization of 
more general second-class constraints systems.

The paper is organized as follows: In the next Section, we start from
discussing a simple, but instructive example of a mechanical system under
second-class holonomic constraints. The one-dimensional reduction of the $O(n)$
non-linear sigma model is discussed, which is equivalent to a mathematical pendulum 
on $n - 1$-dimensional sphere in an $n$-dimensional Euclidean space. Lagrangian 
$\mathcal{L}$ of the system is transformed
on the constraint submanifold of the configuration space into an equivalent
Lagrangian $\mathcal{L}_{*}$ to make explicit the appearance of an underlying gauge 
symmetry. The corresponding Hamiltonian, its constraints structure, and
transformation properties are described in Sect. 3.

This example is analyzed further in Sect. 6 to construct a link with 
the Dirac's quantization method.

In Sects. 4, 5, and 8, the quantization of systems under the
second-class constraints is discussed. 

The Poison bracket of the constraint functions $\mathcal{G}_{a}$ forms a 
symplectic structure in the space of constraint functions. The corresponding 
symplectic basis is suitable for splitting the constraints into canonically conjugate pairs 
($\chi _{A}$,$\Omega _{A}$). In Sect. 4, an algorithm is proposed for constructing
the functions $\chi_{A}$ and $\Omega _{A}$, describing the constraint submanifold 
of second-class constraints systems, for which the
involution relations $\{\chi _{A},\chi _{B}\}=0$, $\{\Omega _{A},\Omega
_{B}\}=0$, and $\{\chi _{A},\Omega _{B}\}=\delta _{AB}$ hold in the strong sense
in an entire neighborhood of any fixed point of the constraint submanifold.

In Sect. 5, properties of the gradient projections into constraint submanifolds described
by equations $\chi _{A}=0$, $\Omega _{A}=0$, and onto their intersection
are discussed. The gradient projection is useful for calculation of the Dirac
bracket and for constructing symplectic basis for the constraint functions. 

The Dirac's quantization of the model of Sect. 2 is performed in Sect. 6 and
the equivalence with the results of Sect. 3 is established. In Appendix A,
the model of a mathematical pendulum on $n-1$-dimensional sphere is embedded 
further into a space of higher dimension to study the constraints 
structure and the associated gauge invariance. The equivalence with the 
results of Sects. 3 and 6 is proved.

In quantum mechanics, systems are described by wave functions or,
equivalently, by density matrices or the Wigner functions. The Wigner
functions are the most closely related to the classical probability
densities in the phase spaces. In Sect. 8, we analyze constraints to be
imposed on the Wigner functions within the framework of the generalized
Hamiltonian dynamics. Such constraints are not unique and depend decisively 
on the way the Wigner functions are extrapolated from the constraint submanifold into
the unconstrained phase space. The main result of this Section consists in
demonstrating the fact that constraints imposed on the Wigner functions can
be taken in a symmetric fashion with regard to permutations 
$\chi_{A} \leftrightarrow \Omega_{A}$, $\chi_{A} \leftrightarrow - \Omega_{A}$
and, more generally, canonical transformations preserving the Poisson brackets.
Those constraints are not symmetric in the Dirac's quantization scheme. We show 
that transition amplitudes are not affected by canonical transformations
mixing the constraint functions.

In Sect. 9, the $O(4)$ non-linear sigma model is discussed as a field theory counterpart
of the mathematical pendulum. We perform quantization in the straightforward way by 
solving the constraints from the outset and demonstrate the equivalence of the 
results with the method based on construction of the gauge-invariant model 
of the $O(4)$ non-linear sigma model. The Lagrange measure for the $O(4)$ non-linear 
sigma model, entering the path integral in the configuration space, is constructed and a 
parameterization is proposed for the pion fields, in terms of which the perturbation theory 
consistent with the mean field (MF) approximation can be developed.
The covariance and the unitarity of the $S$-matrix are demonstrated.

The results are summarized in Sect.10.

\section{Gauged Lagrangian for a spherical pendulum}
\setcounter{equation}{0}

We start from discussing a simple example of a mechanical system under
holonomic constraints: A spherical pendulum on an $n-1$-dimensional
sphere $S^{n-1}$ in $n$-dimensional Euclidean space. The trajectories of 
the system are determined as conditional extremals of the action functional 
$\mathcal{A} = \int{\mathcal{L}dt}$ on the constraint submanifold $\chi = 0$, with 
\begin{equation}
\chi = \ln \phi  \label{CONS}
\end{equation}
where $\phi^2 = \phi^{\alpha} \phi^{\alpha}$ and $\alpha = 1,...,n$.
Lagrangian 
\begin{equation}
\mathcal{L} = \frac{1}{2} {\dot \phi}^{\alpha} {\dot \phi}^{\alpha}
\label{LAGR}
\end{equation}
together with the constraint $\chi = 0$ defines a pointlike massive particle on an $n-1$%
-dimensional sphere $S^{n-1}$ as a mechanical analogue of the $O(n)$ non-linear
sigma model. The constraint $\chi = 0$ implies 
\begin{eqnarray}
\phi^{\alpha}{\dot \phi}^{\alpha} = 0,  \label{CON1} \\
{\dot \phi}^{\alpha}{\dot \phi}^{\alpha} + \phi^{\alpha}{\ddot \phi}%
^{\alpha} = 0,  \label{CON2} \\
...  \nonumber
\end{eqnarray}
The dots stand for higher time derivatives of the constraint function $\chi$%
. Eq.(\ref{CON1}) shows that the radial component of the velocity vanishes.
The second equation shows that the radial component of the acceleration 
equals to the centrifugal force.

The equations of motion can be found using the D'Alembert-Lagrange
variational principle for conditional extremals of the action functionals,
or equivalently, the Euler-Lagrange variational principle for unconditional
extremals with the constraints implemented through the Lagrange's
multipliers method.

Let us substitute $\mathcal{L} \rightarrow \mathcal{L}_1 = \mathcal{L} +
\lambda \chi$ and perform unconditional variations over the Lagrange's
multiplier $\lambda$ and the coordinates $\phi^{\alpha}$. One gets $\chi = 0$ and ${%
\ddot \phi}^{\alpha} = \lambda \phi^{\alpha}/\phi^2$. Multiplying further
the last equation by $\phi^{\alpha}$ and substituting the result into Eq.(%
\ref{CON2}), one gets $\lambda = - {\dot \phi^{\alpha}}{\dot \phi^{\alpha}}$%
. Given that the $\lambda$ is fixed from Eq.(\ref{CON2}), the higher time
derivatives of $\chi$ vanish identically. The radial component of the
acceleration is determined by the constraint, while the tangent component of
the acceleration vanishes. The equations of motion can be presented in the
form 
\begin{equation}
\Delta^{\alpha \beta}(\phi) \frac{d^2}{dt^2} (\phi^{\beta}/{\phi} ) = 0
\label{EQMO}
\end{equation}
where 
\begin{equation}
\Delta^{\alpha \beta}(\phi) = \delta^{\alpha \beta} - \phi ^{\alpha} \phi
^{\beta}/{\phi ^2}.
\end{equation}

The tensor defined above obeys 
\begin{eqnarray}
\phi^{\alpha} \Delta^{\alpha \beta}(\phi) &=& 0, \\
\Delta^{\alpha \beta}(\phi)\Delta^{\beta \gamma}(\phi) &=& \Delta^{\alpha
\gamma}(\phi).
\end{eqnarray}
It is invariant also with respect to dilatation
\begin{equation}
\Delta^{\alpha \beta}(\phi^{\prime}) = \Delta^{\alpha \beta}(\phi),
\end{equation}
for 
\begin{equation}
\phi^{\alpha} \rightarrow \phi^{\prime \alpha} = \exp(\theta)\phi^{\alpha}
\label{TRAN}
\end{equation}
where $\theta$ is an arbitrary parameter.

Eqs.(\ref{EQMO}) tell that the particle moves without tangent acceleration.
In general, the acceleration orthogonal to a constraint submanifold $\chi = 0$
is fixed to keep the particle on it. In our case, the radial
component of the acceleration is determined by the constraint. Eqs.(\ref
{EQMO}) can also be derived using the D'Alembert-Lagrange variational
principle.

On the submanifold $\chi = 0$, the radial components of the velocities are equal
to zero Eq.(\ref{CON1}), so one can replace ${\dot \phi} ^{\alpha}$ by the tangent velocities $%
\Delta^{\alpha \beta}(\phi){\dot \phi}^{\beta}$ in $\mathcal{L}$.

The conditional extremals of the action functional $\mathcal{A}=\int {%
\mathcal{L}dt}$ do not change also if we divide $\mathcal{L}$ by $\phi
^{2}$ ($=1$). The conditional variational problem for Lagrangian 
\begin{equation}
\mathcal{L}_{*}=\frac{1}{2}\Delta ^{\alpha \beta }(\phi ){{\dot{\phi}}%
^{\alpha }{\dot{\phi}}^{\beta }}/{\phi ^{2}}  \label{LAG2}
\end{equation}
is thus completely equivalent to the conditional variational problem we
started with.

The equations of motion (\ref{EQMO}) determine unconditional extremals of
the action functional $\mathcal{A}_{*}=\int {\mathcal{L}_{*}dt}$ on the 
configuration space ${\mathcal M} = (\phi^{\alpha})$. 

{\it The extremals of the action functionals ${}\mathcal{A}$ and ${}\mathcal{A}_{*}$
under the constraint $\chi =0$ coincide.}

$\mathcal{A}{}_{*}$ depends on the spherical coordinates $\phi ^{\alpha }/{%
\phi }$ which lie on an $n-1$-dimensional sphere of a unit radius. Eqs.(\ref
{EQMO}) for unconditional extremals of the ${}\mathcal{A}_{*}$ in the
coordinate space $\phi ^{\alpha }$ coincide, as it should, with the
D'Alembert-Lagrange equations for extremals of $\mathcal{A}_{*}$ in the
space formed by the spherical coordinates under the condition that they belong to an $n-1$%
-dimensional sphere of a unit radius.

It is seen that Lagrangian (\ref{LAG2}) is invariant with respect to
dilatation (\ref{TRAN}) where $\theta$ is an arbitrary function of time. In
the context of the dynamics defined by (\ref{LAG2}) with no constraints
imposed, $\phi$ turns out to be an arbitrary function of time. It can
always be selected to fulfill the constraint $\chi = 0$ or some other
admissible constraint.

{\it The constraint ${\chi} = 0$ can therefore be treated as a
gauge-fixing condition, the function $\phi$ as a gauge degree of freedom,
the ratios $\phi^{\alpha}/{\phi}$ as gauge invariant observables. }

The equations of motion (\ref{EQMO}) are formulated in terms of the gauge
invariant observables.

The above gauge symmetry is defined "on-shell", i.e., for ${\dot \phi}%
^{\alpha}$ treated as a time derivative of ${\phi}^{\alpha}$. In the
tangent bundle $T{\mathcal M} = ({\phi}^{\alpha},{\dot \phi}^{\alpha})$ 
where the coordinates and
their derivatives are independent, $\mathcal{L}_{*}$ is invariant with
respect to a two-parameter set of transformations: 
\begin{eqnarray}
\phi ^{\alpha} &\rightarrow& \phi ^{\prime \alpha} = \exp(\theta) \phi
^{\alpha},  \label{P1} \\
{\dot \phi}^{\alpha} &\rightarrow& {\dot \phi}^{\prime \alpha} = \exp(\theta)%
{\dot \phi}^{\alpha}  \label{P2}
\end{eqnarray}
and 
\begin{eqnarray}
\phi ^{\alpha} &\rightarrow& \phi ^{\prime \alpha} = \phi ^{\alpha},
\label{V1} \\
{\dot \phi}^{\alpha} &\rightarrow& {\dot \phi}^{\prime \alpha} = {\dot \phi}%
^{\alpha} + \epsilon \phi^{\alpha}.  \label{V2}
\end{eqnarray}
The last two equations describe the invariance under variation of the radial
component of the particle velocities. If the ${\dot \phi}^{\alpha}$ is
treated on-shell, then $\epsilon \sim {\dot \theta}$.

A pointlike particle on an $n-1$-dimensional sphere $S^{n-1}$ has therefore underlying
gauge symmetry connected with dilatation of the coordinates $\phi ^{\alpha
} $. Its physical origin is simple: We allow virtual displacements from
the constraint submanifold and treat such displacements as unphysical gauge
degrees of freedom. The constraint equation $\chi =0$ turns thereby into a
gauge-fixing condition. The physical variables are specified by
projections of the coordinates $\phi ^{\alpha }$ onto an $n-1$-dimensional
sphere of a unit radius. Those projections are the spherical coordinates $%
\phi ^{\alpha }/\phi $.

Let us consider the system (\ref{LAG2}) within the generalized Hamiltonian
framework.

\section{The gauged spherical pendulum Hamiltonian}
\setcounter{equation}{0}

Lagrangian (\ref{LAG2}) is defined outside of the constraint submanifold $\chi =
0$ and invariant with respect to the dilatation (\ref{TRAN}). The value $%
\phi$ is an arbitrary function and becomes a gauge degree of freedom. The
equations of motion (\ref{EQMO}) are derived using the constraint $\chi =
0 $. Being formulated, the equations of motion do not depend, however, on the constraint 
anymore and allow an extension to the unconstrained configuration space ${\mathcal M}$.
The invariance under the dilatation, Eq.(\ref{TRAN}), is a consequence
of the two-parameter set of global symmetry transformations, Eqs.(\ref{P1}) - (%
\ref{V2}), of ${\mathcal L}_{*}$ as a function defined on the tangent bundle $T{\mathcal M}$.

One can consider therefore $\mathcal{A}_{*}$ without imposing any constraints
and treat the equation $\chi =0$ as a gauge-fixing condition.

\subsection{Gauged spherical pendulum within generalized Hamiltonian framework}

The canonical momenta corresponding to ${\dot \phi}^{\alpha}$ are defined
by 
\begin{equation}
\pi ^{\alpha} = \frac{\partial \mathcal{L}_{*}}{\partial {\dot \phi}^{\alpha}%
} = \Delta^{\alpha \beta} (\phi) {\dot \phi}^{\beta}/{\phi^2}.  \label{CAM3}
\end{equation}
They satisfy constraints 
\begin{equation}
\pi ^{\alpha} - \Delta^{\alpha \beta} (\phi) \pi ^{ \beta} \approx 0
\label{prim}
\end{equation}
which are equivalent to the one primary constraint: 
\begin{equation}
\Omega = \phi \pi \approx 0.  \label{TPR3}
\end{equation}

The primary Hamiltonian can be obtained with the use of the Legendre
transform: 
\begin{equation}
\mathcal{H} = \frac{1}{2} {\phi ^2} \Delta^{\alpha \beta} (\phi) \pi
^{\alpha}\pi ^{\beta}.  \label{HAM3}
\end{equation}
For $n=3$, $\mathcal{H}$ is proportional to the orbital momentum
squared. The non-vanishing Poisson bracket relations for the canonical variables are
defined by 
\begin{equation}
\{ \phi ^{\alpha} , \pi ^{\beta} \} = \delta ^{\alpha \beta}.  \label{BRAC}
\end{equation}
The constraint function $\Omega$ is stable with respect to the time
evolution: 
\begin{equation}
\{ \Omega, \mathcal{H} \} = 0.  \label{COOH}
\end{equation}
The relations 
\begin{eqnarray}
\{ \phi^{\alpha} , \Omega \} &=& \phi^{\alpha},  \label{TR11} \\
\{ \pi ^{\alpha} , \Omega \} &=& - \pi ^{\alpha}  \label{TR22}
\end{eqnarray}
show that $\Omega$ generates dilatation of $\phi^{\alpha}$ and
$\pi^{\alpha}$. The transformation law for the
canonical coordinates Eq.(\ref{TR11}) is in agreement with Eq.(\ref{P1}) in
its infinitesimal form. The transformation law for the canonical momenta,
which follows from Eqs.(\ref{P1}), (\ref{P2}), and (\ref{CAM3}), 
\begin{equation}
\pi^{\alpha} \rightarrow \pi^{\prime \alpha} = \exp(- \theta)\pi^{\alpha},
\label{TRA3}
\end{equation}
considered in its infinitesimal form, is in agreement with Eq.(\ref{TR22}) either. The
Hamiltonian $\mathcal{H}$ is gauge invariant under the dilatation.

The roles of the gauge-fixing function $\chi$ and the gauge generator $%
\Omega $ are similar. The function $\chi$ is identically in involution with the
Hamiltonian: 
\begin{equation}
\{ \chi, \mathcal{H} \} = 0.  \label{XIHA}
\end{equation}
The Poisson bracket relations 
\begin{eqnarray}
\{ \phi^{\alpha} , \chi \} &=& 0,  \label{TX11} \\
\{ \pi ^{\alpha} , \chi \} &=& - \phi ^{\alpha}  \label{TX22}
\end{eqnarray}
define the one-parameter set of transformations with respect to which $%
\mathcal{H}$ is invariant. The function $\chi$ generates shifts of the
longitudinal component of the canonical momenta. This symmetry is
connected to the invariance of $\mathcal{L}_{*}$ described by Eqs.(\ref
{V1}) and (\ref{V2}) .

The gauge-fixing condition $\chi = 0$ is admissible: 
\begin{equation}
\{ \chi, \Omega \} = 1.
\end{equation}

The equations of motion generated by the primary Hamiltonian look like 
\begin{eqnarray}
{\dot \phi^{\alpha}} = \{ \phi^{\alpha} , \mathcal{H} \} &=& {\phi^2}
\Delta^{\alpha \beta} (\phi) \pi ^{\beta},  \label{DOT3} \\
{\dot \pi ^{\alpha}} = \{ \pi ^{\alpha}, \mathcal{H} \} &=& - \phi ^{\alpha}
\Delta^{\beta \gamma} (\phi) \pi ^{\beta} \pi ^{\gamma}.  \label{DOT4}
\end{eqnarray}

\subsection{Quantization of spherical pendulum}

The quantization of a mathematical pendulum on an $n-1$-dimensional sphere is 
discussed in Ref.\cite{HONG04} where the standard 
Batalin-Fradkin-Tyutin (BFT) algorithm \cite{BATA,BATY} for 
second-class constraints systems is applied. The second-class constraints appear 
if one starts directly from ${\mathcal L}$ and formulate the conditional variational 
problem for $\chi = 0$. By constructing 
auxiliary fields, it is possible to pass over to an equivalent first-class 
constraint system.

In our approach, we start from ${}\mathcal{A}_{*}$ which is gauge
invariant explicitly, so the constraints appear to be of the first class from the
start. One can therefore quantize the pendulum as a gauge-invariant system
without introducing auxiliary fields.

From the point of view of the generalized Hamiltonian framework, 
the gauge-fixing conditions and the gauge generators play similar roles. 
The function $\chi$ does, however, not generate transformations in $T{\mathcal M}$,
so it appears just as a candidate for gauge-fixing function. 
If we pass to the Lagrangian framework, we can verify that
$\chi = 0$ is the gauge-fixing condition indeed.

The system has the gauge invariance described by the generator $%
\Omega$ and the admissible gauge-fixing condition $\chi = 0$. 

{\it The standard
procedure for gauge theories can therefore be applied for quantization of
the spherical pendulum.}

The system is quantized by the algebra mapping 
$(\phi^{\alpha}, \pi^{\alpha}) \to ({\hat \phi}^{\alpha},{\hat \pi}^{\alpha})$ 
and $\{,\}\to -i/{\hbar}[,]$. Consequently, to any symmetrized function in the phase space 
variables one may associate an operator function. The function is symmetrized 
in such a way that quantal image is a hermitian operator. 

The quantum hermitian Hamiltonian has the form 
\begin{equation}
{\hat {\mathcal H}} = \frac{1}{2} \phi \Delta^{\alpha \beta} {\hat \pi}^{\beta} \phi \Delta^{\alpha \gamma} 
{\hat \pi}^{\gamma}. \label{OEH}
\end{equation} 
The vector $i\phi \Delta^{\alpha \beta} {\hat \pi}^{\beta}$ gives the angular part of the gradient operator. 
Although it is not conspicuous, ${\hat {\mathcal H}}$ does not depend on the radial coordinate $\phi$.
The constraint operator can be defined as 
\begin{equation}
{\hat \Omega} = (\phi^{\alpha}{\hat \pi}^{\alpha} + {\hat \pi}^{\alpha}\phi^{\alpha})/2. \label{OEO}
\end{equation} 
It acts only on the radial component of $\phi^{\alpha}$, so the relation 
\begin{equation}
[{\hat \Omega},{\hat {\mathcal H}}] = 0
\end{equation}
holds.

The physical subspace of the Hilbert space is singled out by imposing 
the Dirac's supplementary condition 
\begin{equation}
{\hat \Omega} \Psi =0.  \label{DIRA1}
\end{equation}
This condition implies 
\begin{equation}
\Psi = \phi^{-n/2}\Psi_{1}(\phi^{\alpha}/\phi). 
\end{equation}
The physical information is contained in $\Psi_{1}(\phi^{\alpha}/\phi)$. 

The path integral for the evolution operator becomes 
\begin{equation}
Z=\int {\ \prod_{t}\frac{d^{n}\phi d^{n}\pi}{(2\pi\hbar)^{n-1}}\delta (\chi )\delta (\Omega )}\exp
\left\{ \frac{i}{\hbar}\int {dt(\pi ^{\alpha }{\dot{\phi}}^{\alpha }-{}\mathcal{H})}%
\right\} .  \label{PARQ1}
\end{equation}

Eqs.(\ref{DIRA1}) and (\ref{PARQ1}) solve the quantization problem for a
mathematical pendulum on an $n-1$-dimensional sphere $S^{n-1}$.

It is desirable that the quantization procedure does not destroy the classical 
symmetries which results in having the supplementary condition (\ref{DIRA1}) satisfied 
by the state $\Psi(t)$ for any value of $t$. This feature can, in general, be violated 
either due to complex terms entering the Hamiltonian or by approximations 
adopted for treating the operator eigenvalues. 

Given the initial state wave function $\Psi(0)$ satisfying (\ref{DIRA1}), 
the final wave function $\Psi(t)$ can be found applying the evolution operator 
(\ref{PARQ1}) on $\Psi(0)$. Since ${\hat \Omega}$ commute with the Hamiltonian, the final 
state wave function obeys Eqs.(\ref{DIRA1}).

In the standard canonical frame \cite{MASKAWA}, the first canonical coordinates and momenta 
are identical with the constraint functions. If the Poisson bracket of the Hamiltonian 
with the constraint functions vanishes, the Hamiltonian is independent of the 
first canonical coordinates and momenta. It means that the quantized Hamiltonian is commutative 
with operators associated to the constraint functions. 

{\it By initiating the quantization procedure
in the standard canonical frame, one gets the classical symmetries preserved 
and the Dirac's supplementary condition fulfilled at any time.}

The fact that the dilatation symmetry for the spherical pendulum is preserved on the quantum level is
connected with the fact that the canonical frame we used is simply related to the standard canonical frame.

The integral over the canonical momenta Eq.(\ref{PARQ1}) can be simplified to give 
\begin{equation}
Z=\int {\ \prod_{t}\sqrt{\left( {\partial \chi }/{\partial \phi ^{\alpha }}%
\right) ^{2}}\delta (\chi )d^{n}\phi \exp \left\{ \frac{i}{\hbar}\int {dt{}\mathcal{L}_{*}}%
\right\} }.  \label{MEAS1}
\end{equation}
Lagrange's measure $\sqrt{\left( {\partial \chi }/{\partial \phi ^{\alpha }}\right) ^{2}}\delta (\chi )d^{n}\phi$ 
coincides with the volume element of $S^{n-1}$ sphere. It can be rewritten as an invariant volume of the
configuration space, e.g., in terms of the angular variables $(\varphi_{1}...\varphi _{n-1})$ with the help 
of the induced metric tensor. It is invariant under $O(n)$ rotations and remains the same for all 
functions $\chi $ vanishing at $\phi =1$. For $n=4$, Lagrange's measure 
matches Haar's measure of the group $SU(2)$.

In gauge theories, the evolution operator is independent of the gauge-fixing conditions 
\cite{FADD}. We can insert $\det \{  \chi, \Omega\} = 1$ into the integrand 
of Eq.(\ref{MEAS1}) to bring the measure into the form identical with gauge theories. In the 
path integral, $\chi$ can then be replaced with an arbitrary function. The condition 
(\ref{DIRA1}) does not comprise the constraint $\chi = 0$ also. 

{\it It is remarkable that physical observables depend on half of the number of the second-class 
constraints.}

The main question to be raised here if the quantization method described above is general enough 
or specific only for the spherical pendulum? In Sect. 7 we show that such a method works 
for mechanical systems and field theories under holonomic constraints. The quantization of 
more general second-class constraints systems is discussed in Sect. 8.

\section{Local symplectic basis for second-class constraints functions}
\setcounter{equation}{0}

Two sets of the constraint functions are equivalent if they describe the
same constraint submanifold. The Hamiltonian function admits transformations which 
do not change its value and its first derivatives on the constraint submanifold. 
This allows to make transformations of the constraint and Hamiltonian functions 
without changing the physical content of theory. 


The second-class constraints satisfy $\det \{ \mathcal{G}_{a},\mathcal{G}%
_{b} \} \neq 0$. The Poisson bracket defines therefore a non-degenerate
symplectic linear structure in the vector space of the constraint
functions $\mathcal{G}_{a}$. Indeed, any linear transformation $\mathcal{G}%
^{\prime}_{a} = \mathcal{O}_{ab} \mathcal{G}_{b}$ with matrix $\mathcal{O}%
_{ab}$ depending on the canonical variables transforms accordingly the
Poisson bracket: $\{ \mathcal{G}^{\prime}_{a},\mathcal{G}^{\prime}_{b} \}
\approx \mathcal{O}_{ac}\mathcal{O}_{bd} \{ \mathcal{G}_{c},\mathcal{G}_{d}
\}$. The Poisson bracket plays thereby a role of a skew-scalar product in the
symplectic vector space of the constraint functions. Every symplectic space
has a symplectic basis (see, e.g., \cite{ARNO}), so the constraint functions
can be brought by linear transformations into the form 
\begin{equation}
\{\mathcal{G}_{a},\mathcal{G}_{b}\}\approx \mathcal{I}_{ab}  \label{WEAKCR}
\end{equation}
where 
\begin{equation}
\left\| \mathcal{I}_{ab} \right\| =\left\| 
\begin{array}{ll}
0 & E_{m} \\ 
-E_{m} & 0
\end{array}
\right\|, \label{SYMPL} 
\end{equation}
with $E_{m}$ being the $m\times m$ identity matrix, $\mathcal{I}_{ab}\mathcal{I}%
_{bc}=-\mathcal{\delta }_{ac}$. Using representation $\mathcal{G}_{a} =
(\chi_{A}, \Omega_{A})$, one has 
\begin{eqnarray}
\{ \Omega_{A}, \Omega_{B} \} &\approx& 0,\label{SPLITOO} \\ 
\{ \chi_{A}, \chi_{B} \} &\approx& 0, \label{SPLITCC} \\
\{ \chi_{A}, \Omega_{B} \} &\approx& \delta_{AB}. \label{SPLITCO}
\end{eqnarray}
This basis is not unique. Indeed, there remains a group of symplectic 
transformations $Sp(2m)$, which keeps the Poison bracket of the constraint functions in the
symplectic form.

{\it At any given point of a neighborhood of the constraint submanifold
one can find symplectic basis for second-class constraints
functions in the weak form.}

This result can be strengthened as shown below:

($\mathcal{A}$) If $\xi$ is close enough
to the constrain submanifold, one has $\det \{\chi _{A}(\xi )$,$\Omega _{B}(\xi )\} \neq 0$. 
By continuity there exists a finite neighborhood $\Delta_{\xi }$ of $\xi $, such that 
$\det \{\chi _{A}(\xi ^{\prime })$,$\Omega_{B}(\xi ^{\prime })\}\neq 0\ $ 
$\forall \xi ^{\prime }\in \Delta_{\xi}$. Let us assume that the intersection 
of $\Delta_{\xi}$ with the constraint submanifold is  not empty, i.e., one can find 
$\xi_1 \in \Delta_{\xi}$ such that ${\mathcal G_{a}(\xi_{1})} = 0$. If it is not fulfilled, 
we start from another $\xi$ closer to the constraint submanifold.

Let us chose an equivalent set of the constraint functions $\chi
_{A}\rightarrow \chi _{A}^{\prime }=\{\chi _{A},\Omega _{B}\}^{-1}\chi _{B}$
to ensure $\{\chi _{A}^{\prime },\Omega _{B}\}\approx \delta _{AB}$ in the
region $\Delta _{\xi }$.

($\mathcal{B}$) We replace further $\Omega _{A}\rightarrow \Omega
_{A}^{\prime }=\Omega _{A}-\frac{1}{2}C_{AB}\chi _{B}^{\prime }$ to get
equations 
\begin{equation}
\{\Omega _{A}^{\prime },\Omega _{B}^{\prime }\}\approx \{\Omega _{A},\Omega
_{B}\}-C_{AB}+\frac{1}{4}C_{AD}C_{BF}\{\chi _{D}^{\prime },\chi _{F}^{\prime
}\}\approx 0.
\end{equation}
These equations can be solved for matrix $C_{AB}$ in terms of a power series
of the matrix $\{\chi _{D}^{\prime },\chi _{F}^{\prime }\}$: 
\begin{equation}
C_{AB}=\sum_{k=0}^{\infty }C_{AB}^{[k]}  \label{series}
\end{equation}
where $C_{AB}^{[k]}=-C_{BA}^{[k]}$ and 
\begin{eqnarray*}
C_{AB}^{[0]} &=&\{\Omega _{A},\Omega _{B}\}, \\
C_{AB}^{[k+1]} &=&\frac{1}{4}\sum_{j=0}^{k}C_{AD}^{[k-j]}C_{BF}^{[j]}\{\chi
_{D}^{\prime },\chi _{F}^{\prime }\},
\end{eqnarray*}
with $k=0,1,...$ .

($\mathcal{C}$) The transform $\chi _{A}^{\prime }\rightarrow \chi
_{A}^{\prime \prime }=\{\chi _{A}^{\prime },\Omega _{B}^{\prime }\}^{-1}\chi
_{B}^{\prime }$ brings back the Poisson bracket $\{\chi _{A}^{\prime \prime
},\Omega _{B}^{\prime }\}$ to the diagonal form $\delta _{AB}.$

($\mathcal{D}$) $\,$The last transform looks like $\chi _{A}^{\prime \prime
}\rightarrow \chi _{A}^{\prime \prime \prime }=\chi _{A}^{\prime \prime }-%
\frac{1}{2}\{\chi _{A}^{\prime \prime },\chi _{B}^{\prime \prime }\}\Omega
_{B}^{\prime }.$ It provides $\{\chi _{A}^{\prime \prime \prime },\chi
_{B}^{\prime \prime \prime }\}\approx 0\,$and keeps the relation $\{\chi
_{A}^{\prime \prime \prime },\Omega _{B}^{\prime }\}\approx \delta _{AB}$
unchanged$.$

Now, we remove primes from the notations. As a result of the steps 
($\mathcal{A}$)-($\mathcal{D}$), we obtain weak equations
\begin{equation}
\{ {\mathcal G}_{a}(\zeta),{\mathcal G}_{b}(\zeta)\} \approx \mathcal{I}_{ab}  \label{STROCR}
\end{equation}
$\forall \zeta \in \Delta _{\xi }$. 
It is manifest that Eq.(\ref{STROCR}) is valid
in some neighborhood of any point $\xi_{1}$ of the constraint 
submanifold too. 

{\it The symplectic basis for second-class constraints functions in the weak form 
exists in an entire neighborhood of any given point of the constraint submanifold.}

The existence of the local symplectic basis in the weak form
is on the line with the Darboux's theorem (see, e.g., \cite{RABJAR}) which states that around 
every point $\xi$ in a symplectic space there exists a coordinate 
system in $\Delta_{\xi}$ such that $\xi \in \Delta_{\xi}$ where the symplectic structure takes 
the standard canonical form. The symplectic space can be covered by such coordinate systems.

This is in sharp contrast to the situation in Riemannian geometry where the 
metric at any given point $x$ can always be made Minkowskian, but in any neighborhood 
of $x$ the variance of the Riemannian metric with the Minkowskian metric is, in general,
$\sim \Delta x^2$. In other words, by passing to an inertial coordinate frame one can 
remove gravitation fields at any given point, but not in an entire neighborhood of that point.
The Darboux's theorem states, reversely, that the symplectic 
structure can be made to take the standard canonical form in an entire neighborhood 
$\Delta_{\xi}$ of any point $\xi \in \Delta_{\xi}$. 
In Riemannian spaces, locally means at some given point. In symplectic spaces,
locally means at some given point and in an entire neighborhood of that point.

Locally, all symplectic spaces are indistinguishable. Any submanifold in a symplectic 
space, including any constraint submanifold, is a plane. The possibility of finding 
the standard canonical frame \cite{MASKAWA} illustrates this circumstance.

In the view of this marked dissimilarity, the validity of Eqs.(\ref{STROCR}) in 
a finite domain looks indispensable.

The global symplectic basis exists apparently for $m=1$ as, e.g., in the case 
of the spherical pendulum. The global existence of the basis (\ref{STROCR}) 
has been proved for systems with one primary constraint \cite{MITRA1}, and also, 
assuming that $\det \{\chi _{A},\Omega _{B}\}\neq 0$ holds globally \cite{VYTHE1}.

Admissible transformations on the second-class constraints functions allow to bring 
Eqs.(\ref{STROCR}) into a strong form. The Hamiltonian can also be modified
to convert the Poisson bracket relations with the second-class constraints functions 
into the strong form without changing the dynamics. The arguments given below follow 
closely our discussion of holonomic systems \cite{MFRF}:
 
According to Eqs.(\ref{STROCR}), at any given point of the constraint submanifold 
where $\det \{ \mathcal{G}_{a},\mathcal{G}_{b} \} \neq 0$ one can select symplectic basis
in which the constraint functions satisfy 
\begin{equation}
\{ \mathcal{G}_{a},\mathcal{G}_{b} \} = \mathcal{I}_{ab} + \mathcal{C}%
_{ab\;}^{\;\;c} \mathcal{G}_{c}  \label{POISCO}
\end{equation}
in an entire neighborhood of that point. The first-class Hamiltonian 
$\mathcal{H}$ has relations 
\begin{equation}
\{\mathcal{G}_{a}, \mathcal{H} \} = \mathcal{R}_{a\;}^{\;b} \mathcal{G}_{b}. \label{POISHO}
\end{equation}

The Jacobi identities for $\mathcal{G}_{a}$, $\mathcal{G}_{b}$, and $%
\mathcal{G}_{c}$ and for $\mathcal{G}_{a}$, $\mathcal{G}_{b}$, and $\mathcal{%
H}$ imply 
\begin{eqnarray}
\mathcal{C}_{abc} + \mathcal{C}_{bca} + \mathcal{C}_{cab} &\approx& 0,
\label{RRR} \\
\mathcal{R}_{ab} - \mathcal{R}_{ba} &\approx& 0  \label{ASYM}
\end{eqnarray}
where $\mathcal{C}_{abc} = \mathcal{I}_{cd}\mathcal{C}_{ab\;}^{\;\;d}$ and $%
\mathcal{R}_{ab} = \mathcal{I}_{ac} \mathcal{R}_{a\;}^{\;c}$.

Let us define 
\begin{eqnarray}
\mathcal{G}_{a}^{\prime} &=& \mathcal{G}_{a} + \frac{1}{3} \mathcal{C}%
_{a\;\;}^{\;bc} \mathcal{G}_{b} \mathcal{G}_{c}, \\
\mathcal{H}^{\prime} &=& \mathcal{H} - \frac{1}{2} \mathcal{R}^{ab} \mathcal{%
G}_{a} \mathcal{G}_{b}.
\end{eqnarray}
The constraint functions $\mathcal{G}_{a}$ and $\mathcal{G}_{a}^{\prime}$
coincide on and in vicinity of the constraint submanifold up to the second order
in $\mathcal{G}_{a}$ and generate accordingly identical phase flows on
the submanifold $\mathcal{G}_{a} = 0$. The Hamiltonian functions $\mathcal{H}$
and $\mathcal{H}^{\prime}$ coincide on and in vicinity of the constraint
submanifold up to the second order in $\mathcal{G}_{a}$ and generate on the
submanifold $\mathcal{G}_{a} = 0$ identical Hamiltonian phase flows.

Using Eqs.(\ref{RRR}) and (\ref{ASYM}) one gets 
\begin{eqnarray}
\{\mathcal{G}_{a}^{\prime}, \mathcal{G}_{b}^{\prime} \} &=& \mathcal{I}_{ab}
+ \mathcal{C}_{ab\;\;}^{\;\;cd} \mathcal{G}_{c}^{\prime} \mathcal{G}%
_{d}^{\prime}, \\
\{\mathcal{G}_{a}^{\prime}, \mathcal{H}^{\prime} \} &=& \mathcal{R}%
_{a\;\;}^{\;bc} \mathcal{G}_{b}^{\prime} \mathcal{G}_{c}^{\prime}
\end{eqnarray}
where $\mathcal{C}_{ab\;\;}^{\;\;cd}$ and $\mathcal{R}_{a\;\;}^{\;bc}$ are
new structure functions. The first-order terms in $\mathcal{G}_{a}$ do
not appear in the right sides of these equations. The pair ($\mathcal{G}%
_{a}^{\prime}$,$\mathcal{H}^{\prime}$) describes the same Hamiltonian
dynamics as ($\mathcal{G}_{a}$,$\mathcal{H}$), being at the same time in a
stronger involution around the constraint submanifold.

The above procedure can be repeated to remove the
quadratic terms in $\mathcal{G}_{a}$, cubic terms in $\mathcal{G}%
_{a} $, etc. In general, assuming 
\begin{eqnarray}
\{\mathcal{G}_{a}^{[k]},\mathcal{G}_{b}^{[k]}\} &=&\mathcal{I}_{ab}+\mathcal{%
C}_{ab}^{\;\;c_{1}...c_{k}}\mathcal{G}_{c_{1}}^{[k]}...\mathcal{G}%
_{c_{k}}^{[k]},  \label{GAGB} \\
\{\mathcal{G}_{a}^{[k]},\mathcal{H}^{[k]}\} &=&\mathcal{R}%
_{a\;\;}^{\;b_{1}...b_{k}}\mathcal{G}_{b_{1}}^{[k]}...\mathcal{G}%
_{b_{k}}^{[k]},  \label{GAH}
\end{eqnarray}
we get as a consequence of the Jacobi identities and of the symmetry of
structure functions $\mathcal{C}_{ab\;\;}^{\;\;c_{1}...c_{k}}$ and $\mathcal{%
R}_{a\;\;}^{\;b_{1}...b_{k}}$ with respect to the upper indices: 
\begin{eqnarray}
\mathcal{C}_{abc_{1}...c_{k}}+\mathcal{C}_{bc_{1}a...c_{k}}+\mathcal{C}%
_{c_{1}ab...c_{k}} &\approx &0,  \label{RRRN} \\
\mathcal{R}_{ab_{1}...b_{k}}-\mathcal{R}_{b_{1}a...b_{k}} &\approx &0.
\label{ASYN}
\end{eqnarray}
The next-order constraint functions and the Hamiltonian are given by 
\begin{eqnarray}
\mathcal{G}_{a}^{[k+1]} &=&\mathcal{G}_{a}^{[k]}+\frac{1}{k+2}\mathcal{C}%
_{a}^{\;c_{0}c_{1}...c_{k}}\mathcal{G}_{c_{0}}^{[k]}\mathcal{G}%
_{c_{1}}^{[k]}...\mathcal{G}_{c_{k}}^{[k]}, \;\;\;\;\;\;\;\;  \label{GNEXT} \\
\mathcal{H}^{[k+1]} &=&\mathcal{H}^{[k]}-\frac{1}{k+1}\mathcal{R}%
^{b_{0}b_{1}...b_{k}}\mathcal{G}_{b_{0}}^{[k]}\mathcal{G}_{b_{1}}^{[k]}...%
\mathcal{G}_{b_{k}}^{[k]}. \;\;\;\;\;\;\;\; \label{HNEXT}
\end{eqnarray}
Using Eqs.(\ref{GAGB}) and (\ref{GAH}), one can calculate the next-order
structure functions ${\mathcal C}_{ab\;\;}^{\;\;c_{1}...c_{k+1}}$ and $%
{\mathcal R}_{a\;\;}^{\;b_{1}...b_{k+1}}$ and repeat the procedure. If structure
functions vanish, we shift $k$ by one unit and check Eqs.(\ref{GAGB})
and (\ref{GAH}) again. At each step, the Poisson bracket relations get
closer to the normal form. In the limit $k\rightarrow +\infty $, we obtain 
\begin{eqnarray}
\{{\tilde{\mathcal{G}}}_{a},{\tilde{\mathcal{G}}}_{b}\}&=&{}\mathcal{I}_{ab}, \label{TQ1} \\
\{{\tilde{\mathcal{G}}}_{a},{\tilde{\mathcal{H}}}\}&=&0 \label{TQ2}
\end{eqnarray}
where ${\tilde{\mathcal{G}}}_{a} = \lim_{k \rightarrow \infty} {\mathcal{G}}_{a}^{[k]}$ 
and ${\tilde{\mathcal{H}}} = \lim_{k \rightarrow \infty} {\mathcal{H}}^{[k]}$. 
The matrix $\mathcal{I}_{ab}$ defines splitting of the constraints into two groups (${%
\tilde{\chi}}_{A}$,${\tilde{\Omega}}_{A}$), such that 
\begin{eqnarray}
\{{\tilde{\chi}}_{A},{\tilde{\chi}}_{B}\} &=& 0, \label{PBCC} \\
\{{\tilde{\Omega}}_{A},{\tilde{\Omega}}_{B}\}&=&0, \label{PBOO} \\
\{{\tilde{\chi}}_{A},{\tilde{\Omega}}_{B}\}&=&\delta _{AB}. \label{PBCO}
\end{eqnarray}

The progress made consists in extending
the validity of Eqs.(\ref{WEAKCR}) from one point into its neighborhood 
Eqs.(\ref{STROCR}), and further, in passing from the weak to the strong 
form of the Poisson bracket relations Eqs.(\ref{TQ1}).

{\it The symplectic basis for second-class constraints functions
exists in the strong form in an entire neighborhood of any given point of the constraint submanifold. There 
exits a Hamiltonian function ${\tilde {\mathcal H}}$, describing the same dynamics 
on the constraint submanifold as the initial Hamiltonian function ${\mathcal H}$,
which is identically in involution with the constraint functions.}

An independent geometric-based construction of 
${\tilde{\mathcal{G}}}_{a}$ and ${\tilde{\mathcal{H}}}$ is given in Sect. 5.

A similar local basis exists for first-class constraints \cite{HENN,SCHO49}. 
The arguments of Refs.\cite{HENN,SCHO49} apply to second-class constraints under 
special restrictions which are discussed in Appendix B.

For systems of pointlike particles under holonomic constrains, Eqs.(\ref{POISCO}) hold globally, 
so ${\tilde{\mathcal{G}}}_{a}$ and ${\tilde{\mathcal{H}}}$ exist globally also. Furthermore,
${\tilde{\Omega}}_{A}$ is linear in the canonical 
momenta, ${\tilde{\chi}}_{B}$ does not depend on the canonical momenta, and
${\tilde{\mathcal{H}}}$ splits into a sum of a kinetic energy term quadratic in the canonical 
momenta and a potential energy term depending on the canonical coordinates \cite{MFRF}.

\section{Gradient projection}
\setcounter{equation}{0}

The concept of the gradient projection is useful for applications. It defines 
functions $\xi _{s}(\xi )$ which project an arbitrary point $\xi $ of the phase 
space onto a submanifold $\mathcal{G}_{a} = 0$ ($a = 1,...,2m$) of
the phase space along phase flows generated by the constraint functions $\mathcal{G}_{a}$. 

\subsection{Full gradient projection}

Let $\mathcal{G}_{a} = 0$ be second-class constraints, 
$\det \{ \mathcal{G}_{a},\mathcal{G}_{b}\} \neq 0$, $\xi^{i}$ ($i = 1,...,2n$%
) are canonical variables. In vicinity of the submanifold $\mathcal{G}_{a} = 0$,
the projections can be constructed explicitly. Near the
constraint submanifold, one can write $\xi_{s}(\xi) = \xi + \{ \xi, \mathcal{G}%
_{a}\} \lambda_{a}$. The small parameters $\lambda_{a}$ are determined by
requiring $\mathcal{G}_{a}(\xi_{s}(\xi)) = 0$ to the first order in $%
\mathcal{G}_{a}$. We get 
\begin{equation}
\xi_{s}(\xi) = \xi - \{ \xi, \mathcal{G}_{a}\} ||\{ \mathcal{G}_{a}, 
\mathcal{G}_{b} \} ||^{-1} \mathcal{G}_{b}.  \label{PROJXO}
\end{equation}

It is seen that $\{ \mathcal{G}_{a}, \xi_{s}(\xi) \} = 0$, consequently $\{ \mathcal{G}%
_{a}, f(\xi_{s}(\xi))\} = 0$ for any function $f$. This is natural, since
$\mathcal{G}_{a}$ generate phase flows along which the projections $%
\xi_{s}(\xi)$ have been constructed.

The reciprocal statement is also true: If $\{ \mathcal{G}_{a}, f \} = 0$ for
all $\mathcal{G}_{a}$, then $f = f(\xi_{s}(\xi))$. Indeed, the
coordinates on the constraint submanifold can be parameterized by $\xi_{s}$.
The coordinates describing shifts from the constraint submanifold can be
parameterized by $\mathcal{G}_{a}$. The functions $f$ can in general be
written as $f = f(\xi_{s},\mathcal{G}_{a})$. If the Poisson brackets of $f$ 
with all $\mathcal{G}_{a}$ vanish, $f$ depends on $\xi_{s}$ only.

This can be summarized by 
\begin{equation}
\{ \mathcal{G}_{a}, f \} = 0 \leftrightarrow f = f(\xi_{s}(\xi)).  \label{XO}
\end{equation}

Beyond the lowest order in $\mathcal{G}_{a}$, the operation is unique provided the 
phase flows commute. This is always the case for the constraint functions taken to 
accomplish (\ref{TQ1}) in a finite neighborhood of the constraint submanifold.

\subsection{Partial gradient projection}

Let $\mathcal{G}_{a}$ split into $\chi_{A}$ and $\Omega_{A}$. We wish to
construct functions $\xi_{u}(\xi)$ which project an arbitrary point $\xi$ of
the phase space onto the gauge-fixing surface $\chi_{A} = 0$ with the use
of the constraint functions $\Omega_{A}$ associated to gauge transformations. 
In vicinity of the submanifold, one can
write $\xi_{u}(\xi) = \xi + \{ \xi, {\Omega}_{A}\} \lambda_{A}$. To the
first order in ${\chi}_{A}$, the parameters $\lambda_{A}$ can be found
from equation ${\chi}_{A}(\xi_{u}(\xi)) = 0$: 
\begin{equation}
\xi_{u}(\xi) = \xi - \{ \xi, {\Omega}_{A}\} ||\{ {\chi}_{A}, {\Omega}_{B} \}
||^{-1} {\chi}_{B}.  \label{PROJXX}
\end{equation}

The projection is made along the phase flows of $\Omega_{A}$, so $\{ \Omega_{A},
f(\xi_{u}(\xi)) \} = 0$ for any function $f$.

The reverse statement is also true. We write $f = f(\xi_{u}, \chi_{A})$ and
conclude from $\{ \Omega_{A}, f \} = 0$ that the dependence on $\chi_{A}$
drops out.

It can be summarized as follows: 
\begin{equation}
\{ {\Omega}_{A}, f \} = 0 \leftrightarrow f = f(\xi_{u}(\xi)).  \label{XX}
\end{equation}

The second partial gradient projection can be made onto the submanifold $%
\Omega _{A}=0$ with the use of the constraint functions $\chi _{A}$. The
result is similar to Eq.(\ref{PROJXX}) 
\begin{equation}
\xi _{v}(\xi )=\xi +\{\xi ,{\chi }_{A}\}||\{{\chi }_{A},{\Omega }%
_{B}\}||^{-1}{\Omega }_{B}.  \label{PROJOO}
\end{equation}

The relation $\{ \chi_{A}, f(\xi_{v}(\xi)) \} = 0$ is valid for any function 
$f$. Furthermore, 
\begin{equation}
\{ \chi_{A}, f \} = 0 \leftrightarrow f = f(\xi_{v}(\xi)).  \label{OO}
\end{equation}

Combining the partial projections, e.g., $\xi _{s}(\xi )=\xi _{v}(\xi
_{u}(\xi ))$, one gets the full gradient projection. To the first order in
$\mathcal{G}_{a}$, the order in which the partial projections are applied does not
matter, so $\xi _{v}(\xi _{u}(\xi ))=\xi _{u}(\xi _{v}(\xi ))$. The full
gradient projection constructed in this way coincides with that given by Eq.(\ref{PROJXO}).

\subsection{Example: Gauged spherical pendulum}

The constraint function $\Omega $ can be used to bring the vector $\phi ^{\alpha }$
onto a sphere of a unit radius $\phi =1$ ($\chi =0$). Such a transformation
has a meaning of a gradient projection. The functions $\phi _{u}^{\alpha }$
and $\pi _{u}^{\alpha }$ can be constructed in terms of the variables $\phi
^{\alpha }$ and $\pi _{\alpha }$: 
\begin{eqnarray}
\phi _{u}^{\alpha } &=&\exp (\theta )\phi ^{\alpha },  \label{NEW3} \\
\pi _{u}^{\alpha } &=&\exp (-\theta )\pi ^{\alpha }.  \label{NEW4}
\end{eqnarray}
The condition $\chi (\phi _{u},\pi _{u})=0$ gives $\exp (\theta )=1/\phi $.

The Poisson bracket relations for the projected variables (\ref{NEW3}) and (\ref{NEW4}%
) can be found to be 
\begin{eqnarray}
\{\phi _{u}^{\alpha },\phi _{u}^{\beta }\} &=&0,  \label{QQ3} \\
\{\phi _{u}^{\alpha },\pi _{u}^{\beta }\} &=&\Delta _{u}^{\alpha \beta
}(\phi _{u}),  \label{QP3} \\
\{\pi _{u}^{\alpha },\pi _{u}^{\beta }\} &=&\phi _{u}^{\beta }\pi
_{u}^{\alpha }-\phi _{u}^{\alpha }\pi _{u}^{\beta }  \label{PP3}
\end{eqnarray}
where $\Delta _{u}^{\alpha \beta }(\phi _{u})=\delta ^{\alpha \beta }-\phi
_{u}^{\alpha }\phi _{u}^{\beta }$.

The following properties are worthy of mention:

(i) The Poisson bracket of the projected canonical variables coincides with
the Dirac bracket associated to the constraints $\chi = 0$ and $\Omega = 0$
on the submanifold $\chi = 0$.

The Dirac bracket for the canonical variables can be calculated using Eq.(%
\ref{DIBRBR}) to give 
\begin{eqnarray}
\{ \phi ^{\alpha}, \phi ^{\beta} \}_{D} &=& 0,  \label{DQQ5} \\
\{ \phi ^{\alpha}, \pi ^{\beta} \}_{D} &=& \Delta^{\alpha \beta}(\phi),
\label{DQP5} \\
\{ \pi ^{\alpha}, \pi ^{\beta} \}_{D} &=& ( \phi ^{\beta } \pi ^{\alpha} -
\phi ^{\alpha} \pi ^{\beta })/{\phi^2}.  \label{DPP5}
\end{eqnarray}
The right sides of Eqs.(\ref{QQ3}) - (\ref{PP3}) are reproduced at $\chi = 0$%
.

(ii) The relations (\ref{QQ3}) - (\ref{PP3}) define a Poisson algebra in the space
of functions $f(\phi _{u}^{\alpha },\pi _{u}^{\alpha})$ depending on the projected 
canonical variables, so they can be used to generate consistently a Hamiltonian 
phase flow on the constraint submanifold $\chi = 0$.

(iii) The Hamiltonian function 
\begin{equation}
\mathcal{H}=\frac{1}{2}\Delta _{u}^{\alpha \beta }({\phi }_{u})\pi
_{u}^{\alpha }{\pi }_{u}^{\beta }
\end{equation}
coincides with Eq.(\ref{HAM3}): $\mathcal{H}$$=\mathcal{H}({\phi }%
_{u}^{\alpha },{\pi }_{u}^{\alpha })$$=\mathcal{H}({\phi }^{\alpha },{\pi }%
^{\alpha })$. The Hamiltonian (\ref{HAM3}) is thus the function of the
projected variables $\xi _{u}(\xi )$. It can be defined first on the
submanifold $\chi =0$ and then extended to the unconstrained phase space using the
gradient projection parallel to the phase flow associated to $\Omega 
$. The relation $\{\Omega ,\mathcal{H}\}=0$, Eq.(\ref{COOH}), is the necessary
and sufficient condition (see Eq.(\ref{XX})) for $\mathcal{H}$ to be a function of
a fewer number of variables $\mathcal{H}=\mathcal{H}({\phi }_{u}^{\alpha },{%
\pi }_{u}^{\alpha })$.

Let us consider the gradient projection onto the submanifold $\Omega =0$ using
the constraint function $\chi $. The constraint $\chi =0$ is responsible for
shifts of the longitudinal component of the canonical momenta Eqs.(\ref{TX11}%
) and (\ref{TX22}). The functions $\phi _{v}^{\alpha }$ and $\pi
_{v}^{\alpha }$ have the form 
\begin{eqnarray}
\phi _{v}^{\alpha } &=&\phi ^{\alpha },  \label{NEW5} \\
\pi _{v}^{\alpha } &=&\pi ^{\alpha }-\phi ^{\alpha }\phi \pi /\phi ^{2}.
\label{NEW6}
\end{eqnarray}
Equation $\Omega (\phi _{v},\pi _{v})=0$ is fulfilled identically.

The Poisson bracket relations for the projected variables (\ref{NEW5}) and (\ref{NEW6}%
) can be found to be 
\begin{eqnarray}
\{\phi _{v}^{\alpha },\phi _{v}^{\beta }\} &=&0,  \label{QQ5} \\
\{\phi _{v}^{\alpha },\pi _{v}^{\beta }\} &=&\Delta ^{\alpha \beta }(\phi
_{v}),  \label{QP5} \\
\{\pi _{v}^{\alpha },\pi _{v}^{\beta }\} &=&(\phi _{v}^{\beta }\pi
_{v}^{\alpha }-\phi _{v}^{\alpha }\pi _{v}^{\beta })/{\phi _{v}^{2}}.
\label{PP5}
\end{eqnarray}
One can see again:

(i) The Poisson bracket of the projected variables coincides with the Dirac
bracket Eqs.(\ref{DQQ5}) - (\ref{DPP5}) associated to the constraints $%
\chi =0$ and $\Omega =0$ on the submanifold $\Omega =0$.

(ii) The Poisson bracket relations (\ref{QQ5}) - (\ref{PP5}) are closed and
define thereby a Poisson algebra
in the space of functions depending on the projected canonical 
variables $(\phi _{v}^{\alpha },\pi _{v}^{\beta })$.

(iii) The Hamiltonian function 
\begin{equation}
\mathcal{H}=\frac{1}{2}\phi _{v}^{2}\Delta ^{\alpha \beta }({\phi }_{v})\pi
_{v}^{\alpha }{\pi }_{v}^{\beta }.  \label{HA98}
\end{equation}
coincides with Eq.(\ref{HAM3}): $\mathcal{H}=\mathcal{H}({\phi }_{v}^{\alpha
},{\pi }_{v}^{\alpha })=\mathcal{H}({\phi }^{\alpha },{\pi }^{\alpha })$.
$\mathcal{H}$ given by Eq.(\ref{HAM3}) is the function of the gradient
variables $\xi _{v}(\xi )$. The relation $\{\chi ,\mathcal{H}\}=0$, Eq.(\ref
{XIHA}), is the necessary and sufficient condition to present the Hamiltonian
function as a function the projected variables: $\mathcal{H}=\mathcal{H}%
({\phi }_{v}^{\alpha },{\pi }_{v}^{\alpha })$.

It is clear that the $\mathcal{H}$ is defined finally on the intersection of
the submanifolds $\chi =0$ and $\Omega =0$, being thus a function of the $\xi
_{s}(\xi )$. Eq.(\ref{HAM3}) represents its extension to the unconstrained phase
space using the full gradient projection.

The statements (i) - (iii) are of the general validity for gradient
projections. The statement (iii) has been proved as such above, the other
two ones are proved below.

\subsection{Dirac bracket calculated by gradient projection}

The phase flow associated to a function $g=g(\xi )$ defined on the submanifold 
$\mathcal{G}_{a}=0$ has an ambiguity since one can add to $g=g(\xi )$ a
linear combination of the constraints $\lambda _{a}\mathcal{G}_{a}$ where $%
\lambda _{a}$ are undetermined parameters. The phase flow $L_{g}[.]$ applied to a
function $f=f(\xi )$ suffers from this ambiguity also: 
\begin{equation}
L_{g}[f]=\{f,g\}+\lambda _{a}\{f,\mathcal{G}_{a}\}.  \label{FLOW}
\end{equation}
The submanifold $\mathcal{G}_{a}=0$ should, however, be invariant, i.e., $L_{g}[%
\mathcal{G}_{a}]=0$. This equation allows to find $\lambda _{a}=\lambda
_{a}(\xi )$. Substituting $\lambda _{a}$ into Eq.(\ref{FLOW}), one gets the
Dirac bracket 
\begin{eqnarray}
L_{g}[f] &=&\{f,g\}-\{f,\mathcal{G}_{a}\}||\{\mathcal{G}_{a},\mathcal{G}%
_{b}\}||^{-1}\{\mathcal{G}_{b},g\}  \nonumber \\
&=&\{f,g\}_{D}  \label{DIRACBRACKET}
\end{eqnarray}
where $f$, $g$, and $\mathcal{G}_{a}$ are functions of $\xi $. The Dirac
bracket defines a phase flow generated by a function $g=g(\xi )$ within
the constraint submanifold $\mathcal{G}_{a}=0$.

Using Eq.(\ref{PROJXO}), one finds at ${{\mathcal G}_{a} =0}$
\begin{eqnarray}
\{f(\xi ),g(\xi )\}_{D} &=&\{f(\xi ),g(\xi _{s}(\xi ))\}=\{f(\xi _{s}(\xi
)),g(\xi )\}  \nonumber \\
&=&\{f(\xi _{s}(\xi )),g(\xi _{s}(\xi ))\}.  \label{DIBRXO}
\end{eqnarray}
This is the analogue of the statement (i) made in the previous subsection for the
full gradient projection. The gradient projection can therefore be used
to calculate the Dirac bracket.

There is an analogue of this statement for the partial gradient projections also. Let us
suppose that the second-class constraints $\mathcal{G}_{a}=0$ split into the canonical 
pairs: $\chi _{A}=0$ and $\Omega _{A}=0$ such that $\{\chi _{A},\chi
_{B}\}=0$, $\{\Omega _{A},\Omega _{B}\}=0$, and $\det ||\{\chi _{A},\Omega
_{B}\}||\ne 0$. The Dirac bracket becomes 
\begin{eqnarray}
\{f,g\}_{D} &=&\{f,g\}  \label{DIBRBR} \\
&+&\{f,\chi _{A}\}||\{\chi _{A},\Omega _{B}\}||^{-1}\{\Omega _{B},g\} 
\nonumber \\
&-&\{f,\Omega _{A}\}||\{\chi _{A},\Omega _{B}\}||^{-1}\{\chi _{B},g\}. 
\nonumber
\end{eqnarray}
Let $\xi _{u}(\xi )$ and $\xi _{v}(\xi )$ be partial gradient projections
such that $\chi _{A}(\xi _{u}(\xi ))=0$ and $\Omega _{A}(\xi _{v}(\xi ))=0$
identically (cf. Eqs.(\ref{PROJXX}) and (\ref{PROJOO})). If we replace
arguments of the functions $f$ and $g$ with $\xi _{u}(\xi )$ or $\xi
_{v}(\xi )$, the last two terms vanish due to (\ref{XX}) or (\ref{OO}%
), respectively. The Poisson bracket for the projected variables then
coincides with the Dirac bracket for the canonical variables $\xi $
constrained to the submanifold $\chi _{A}=0$ or $\Omega _{A}=0$: 
\begin{eqnarray}
\{f(\xi _{u}(\xi )),g(\xi _{u}(\xi))\}_{D} &=&\{f(\xi _{u}(\xi )),g(\xi _{u}(\xi))\}, \;\;\;\;\; \label{11} \\
\{f(\xi _{v}(\xi )),g(\xi _{v}(\xi))\}_{D} &=&\{f(\xi _{v}(\xi )),g(\xi _{v}(\xi))\}. \;\;\;\;\; \label{22}
\end{eqnarray}
The arguments of the functions represent the partial
gradient projections like in Eqs.(\ref{QQ3}) - (\ref{PP3}) and (\ref{QQ5}) - (\ref
{PP5}).

Eqs.(\ref{11}) and (\ref{22}) are sufficient to calculate the Dirac
bracket given that the partial gradient projections are constructed.

This completes the proof of the statement (i) from the previous subsection,
extended to arbitrary Hamiltonian systems.

Turning to the point (ii), it is sufficient to notice that the Poisson
bracket, e.g., $\{\xi _{s}^{i},\xi _{s}^{j}\}$ determines a variation of the $%
\xi _{s}^{i}$ along the submanifold $\mathcal{G}_{a}=0$. This submanifold is
parameterized with the $\xi _{s}$. The Poisson bracket is thus a function of
the $\xi _{s}$ again. The involution relations for $\{\xi _{s}^{i},\xi
_{s}^{j}\}$ define therefore an algebra. Similar arguments
apply for the partial gradient projections. Eqs.(\ref{QQ3}) - (\ref{PP3})
and (\ref{QQ5}) - (\ref{PP5}) represent therefore an illustration of this
statement.

\subsection{Constraint functions ${\tilde {\mathcal G}}_{a}$ constructed by gradient projection}

The statements of Sect. 4 can be proved using the gradient 
projection method.

The vector fields 
\begin{equation}
{I}^{ij}\frac{\partial {\mathcal G}_{a}}{\partial \xi^{j}} \label{VF}
\end{equation}
determine phase flows associated to the constraint functions ${\mathcal G}_{a}$. These 
fields are non-singular, i.e., do not vanish in a neighborhood of the constraint submanifold 
${\mathcal F} = \{\xi :{\mathcal G}_{a}(\xi) = 0 \; \forall a\}$.
The opposite would mean $\exists a$ such that (\ref{VF}) vanish at some point 
$\xi \in {\mathcal F}$. It follows then that 
$\{ {\mathcal G}_{a},{\mathcal G}_{b} \} = 0$ $\forall b$. 
This is in contradiction with 
\begin{equation}
\det \{ {\mathcal G}_{a},{\mathcal G}_{b} \} \ne 0 \label{NONGEN}
\end{equation} 
which holds, by assumption, 
everywhere on ${\mathcal F}$ and, by continuity, in a neighborhood of ${\mathcal F}$.
In Eq.(\ref{VF}), ${I}^{ij} = - {I}_{ij}$ where
\begin{equation}
\left\| {I}_{ij} \right\| =\left\| 
\begin{array}{ll}
0 & E_{n} \\ 
-E_{n} & 0
\end{array}
\right\|, \label{SYMPLFULL} 
\end{equation}
with $E_{n}$ being the $n\times n$ identity matrix (cf. Eq.(\ref{SYMPL})).
In what follows, we denote phase flows (\ref{VF}) briefly as $Id{\mathcal G}_{a}$. 

Let ${\mathcal F}_{a} = \{\xi :{\mathcal G}_{a}(\xi) = 0 \}$. 
The condition $Id{\mathcal G}_{a}(\xi) \neq 0$ $\forall \xi \in {\mathcal F}_{a}$ 
is stronger than the one mentioned above. It looks evident, since any constraint 
function ${\mathcal G}_{a}(\xi)$ in a neighborhood of ${\mathcal F}_{a}$ can be redefined 
to assign the gradient a definite direction. 

${\mathcal F}_{a}$ is a subspace of the dimension $2n - 1$. The intersection of all ${\mathcal F}_{a}$
gives ${\mathcal F}$. The tangent space to ${\mathcal F}_{a}$ at a point $\xi$, denoted 
as $T_{\xi}{\mathcal F}_{a}$, is skew-orthogonal to $Id{\mathcal G}_{a}(\xi)$. Indeed, 
if $d{\mathcal G}_{a} = 0$ then $d\xi \in T_{\xi}{\mathcal F}_{a}$, and we obtain
\begin{equation}
{I}^{ij}\frac{\partial {\mathcal G}_{a}}{\partial \xi^{j}}{I}_{ik}d\xi^{k} 
= -\frac{\partial {\mathcal
G}_{a}}{\partial \xi^{i}}d\xi^{i} = 0. \label{ORTO}
\end{equation}
The space $T_{\xi}{\mathcal F}_{a}$ has the dimension $2n - 1$.
Among the vectors of $T_{\xi}{\mathcal F}_{a}$ one can find $Id{\mathcal G}_{a}(\xi)$,
since $Id{\mathcal G}_{a}(\xi)$ is skew-orthogonal to itself.
$T_{\xi}{\mathcal F}_{a}$ is therefore a skew-orthogonal complement of 
$Id{\mathcal G}_{a}(\xi)$. 

One can find $b$ such that $\{ {\mathcal G}_{a}(\xi),{\mathcal G}_{b}(\xi) \} \ne 0$.
As discussed above,
\begin{eqnarray}
Id{\mathcal G}_{a}(\xi) &\in& T_{\xi}{\mathcal F}_{a}, \label{Aa} \\
Id{\mathcal G}_{b}(\xi) &\in& T_{\xi}{\mathcal F}_{b}, \label{Bb}
\end{eqnarray}
and furthermore,
\begin{eqnarray}
Id{\mathcal G}_{a}(\xi) &\notin& T_{\xi}{\mathcal F}_{b}, \label{AA} \\
Id{\mathcal G}_{b}(\xi) &\notin& T_{\xi}{\mathcal F}_{a}. \label{BB}
\end{eqnarray}
From the other side,
any vector $d\xi \in T_{\xi}{\mathcal F}_{a} \cap T_{\xi}{\mathcal F}_{b}
= T_{\xi}({\mathcal F}_{a} \cap {\mathcal F}_{b})$ is skew-orthogonal
to $Id{\mathcal G}_{a}(\xi)$ and $Id{\mathcal G}_{b}(\xi)$. 
As a consequence of Eqs.(\ref{AA}) and (\ref{BB}), one gets
\begin{eqnarray}
Id{\mathcal G}_{a}(\xi) \notin T_{\xi}({\mathcal F}_{a} &\cap& {\mathcal F}_{b}), \label{A7} \\
Id{\mathcal G}_{b}(\xi) \notin T_{\xi}({\mathcal F}_{a} &\cap& {\mathcal F}_{b}). \label{A8}
\end{eqnarray}
The subspace ${\mathcal F}_{a} \cap {\mathcal F}_{b}$ and, respectively,
$T_{\xi}({\mathcal F}_{a} \cap {\mathcal F}_{b})$ have the dimension $2n - 2$.
The vectors $Id{\mathcal G}_{a}(\xi)$ and $Id{\mathcal G}_{b}(\xi)$
are linearly independent and form a two-dimensional space 
${\mathcal D}_{ab}(\xi)$ which is a skew-orthogonal complement of 
$T_{\xi}({\mathcal F}_{a} \cap {\mathcal F}_{b})$ such that
${\mathcal D}_{ab}(\xi) \cap T_{\xi}({\mathcal F}_{a} \cap {\mathcal F}_{b}) = \emptyset$.

Let us consider motion of a particle with a Hamiltonian function 
${\mathcal G}_{b}$. In virtue of Eq.(\ref{BB}), the phase flow 
$Id{\mathcal G}_{b}$ does not lie in the submanifold ${\mathcal F}_{a}$ entirely 
and therefore crosses it. Let $t_{a}(\zeta)$ be time needed
for particle located at $\zeta \notin {\mathcal F}_{a}$ in a neighborhood 
of $\xi \in {\mathcal F}_{a}$ to cross ${\mathcal F}_{a}$ at some point $\eta \ne \xi$.
The equations of motion look like 
\[
\frac{d{\zeta}}{dt_a} = Id{\mathcal G}_{b}(\zeta).
\]
The derivative of $t_{a}(\zeta)$ along the phase flow $Id{\mathcal G}_{b}(\zeta)$ 
with respect to time is, by definition, equal to unity:
\begin{equation}
\{t_{a}(\zeta),{\mathcal G}_{b}(\zeta)\} = 1.  \label{TIME}
\end{equation}
One may interpret, equivalently, $-{\mathcal G}_{b}(\zeta)$ as a time needed to cross
the submanifold ${\mathcal F}_{b}$ by a particle located at $\zeta \notin {\mathcal F}_{b}$. 
The motion of such a particle is described by a Hamiltonian function $t_{a}(\zeta)$. 

The function $t_{a}(\zeta)$ vanishes for $\zeta = \eta \in {\mathcal F}_{a}$. At any 
point of ${\mathcal F}_{a}$, $d{\mathcal G}_{a}$ 
and $dt_{a}$ vanish for $d\eta \in T_{\eta}{\mathcal F}_{a}$. There exists 
only one $d\eta \notin T_{\eta}{\mathcal F}_{a}$ such that
$d{\mathcal G}_{a} \ne 0$ and $dt_{a} \ne 0$. It means that $\forall d\eta$ 
$d{\mathcal G}_{a}$ is proportional to $dt_{a}$ and $Id{\mathcal G}_{a}$ is 
in turn proportional to $Idt_{a}$.

The first canonical pair ${\tilde {\mathcal G}}_{a} = t_{a}$ and 
${\tilde {\mathcal G}}_{b} = {\mathcal G}_{b}$ is thus constructed. 

Let us consider the full gradient projection $\xi_{1}(\xi)$ onto the submanifold 
${\mathcal F}_{a} \cap {\mathcal F}_{b}$, using 
the constraint functions ${\tilde {\mathcal G}}_{a}$ and 
${\tilde {\mathcal G}}_{b}$. One gets ${\tilde {\mathcal G}}_{a}(\xi_{1}(\xi)) \equiv 0$, 
${\tilde {\mathcal G}}_{b}(\xi_{1}(\xi)) \equiv 0$, whereas the equations
${\mathcal G}_{c}(\xi_{1}(\xi)) = 0$ for $c \neq a,b$ are significant to determine the location 
of the constraint submanifold ${\mathcal F}$, owing to shifts along the phase flows $Id{\tilde {\mathcal G}}_{a}$
and $Id{\tilde {\mathcal G}}_{b}$. A complete set of equations for ${\mathcal F}$ 
can be taken to be
\begin{eqnarray}
{\tilde {\mathcal G}}_{a}(\xi) &=& 0, \\ \label{w1}
{\tilde {\mathcal G}}_{b}(\xi) &=& 0, \\ \label{w2}
{\mathcal G}_{c}(\xi_{1}(\xi)) &=& 0     \label{w3}
\end{eqnarray}
for $c \neq a,b$. In virtue of Eq.(\ref{XO}), 
\begin{eqnarray}
\{ {\tilde {\mathcal G}}_{a}(\xi),{\mathcal G}_{c}(\xi_{1}(\xi)) \} &=& 0, \\
\{ {\tilde {\mathcal G}}_{b}(\xi),{\mathcal G}_{c}(\xi_{1}(\xi)) \} &=& 0.
\end{eqnarray}
Eqs.(\ref{w1})-(\ref{w3}) determine ${\mathcal F}$
uniquely, so the determinant of the Poisson bracket relations between the $2m$ 
constraint functions is not zero. The functions 
${\tilde {\mathcal G}}_{a}$ and ${\tilde {\mathcal G}}_{b}$ have the vanishing 
Poisson brackets with the rest ones, so
\begin{equation}
\det \{ {\mathcal G}_{c}(\xi_{1}(\xi)),{\mathcal G}_{d}(\xi_{1}(\xi))\} \ne 0
\end{equation}
where $c,d$ take $2m - 2$ values ($c,d \ne a,b$). 

In the remaining set of the constraints, one can find $c,d$ such that 
$\{ {\mathcal G}_{c}(\xi_{1}(\xi)),{\mathcal G}_{d}(\xi_{1}(\xi)) \} \ne 0$ and
repeat the arguments we used earlier. The analogue of Eq.(\ref{TIME}) looks like
\begin{equation}
\{t_{c}(\zeta),{\mathcal G}_{d}(\xi_{1}(\zeta))\} = 1.  \label{TIME2}
\end{equation}
The Poisson brackets of the left side of this equation with ${\tilde {\mathcal G}}_{a}(\zeta)$ and 
${\tilde {\mathcal G}}_{a}(\zeta)$ vanish. The Jacoby identity yields
\begin{eqnarray}
\{ \{t_{c}(\zeta),{\tilde {\mathcal G}}_{a}(\zeta) \}, {\mathcal G}_{d}(\xi_{1}(\zeta))\} &=& 0,  \\ \label{TIME3}
\{ \{t_{c}(\zeta),{\tilde {\mathcal G}}_{b}(\zeta) \}, {\mathcal G}_{d}(\xi_{1}(\zeta))\} &=& 0.     \label{TIME4}
\end{eqnarray}
The Poisson brackets of $t_{c}(\zeta)$ with 
${\tilde {\mathcal G}}_{a}(\zeta)$ and ${\tilde {\mathcal G}}_{b}(\zeta)$ remain therefore 
constant along the phase flow $Id{\mathcal G}_{d}(\xi_{1}(\zeta))$. At the submanifold 
${\mathcal F}_{c}^{\prime} = \{ \zeta : {\mathcal G}_{c}(\xi_{1}(\zeta)) = 0 \}$, 
$Idt_{c}(\zeta)$ is proportional to $Id{\mathcal G}_{c}(\xi_{1}(\zeta))$. Those brackets 
vanish at ${\mathcal F}_{c}^{\prime}$, and furthermore, vanish for 
$\zeta \notin {\mathcal F}_{c}^{\prime}$. 
Eq.(\ref{XO}) suggests then $t_{c}(\zeta) = t_{c}(\xi_{1}(\zeta))$.

The second canonical pair ${\tilde {\mathcal G}}_{c}(\zeta) = t_{c}(\xi_{1}(\zeta))$ and 
${\tilde {\mathcal G}}_{d}(\zeta) = {\mathcal G}_{d}(\xi_{1}(\zeta))$ is thus constructed. 

The proof can be completed by induction. We consider the full gradient 
projection $\xi_{2}(\zeta)$ onto the submanifold ${\mathcal F}_{a} \cap {\mathcal F}_{b} \cap {\mathcal F}_{c} \cap {\mathcal F}_{d}$
along the commutative phase flows generated by 
${\tilde {\mathcal G}}_{a}(\zeta)$, 
${\tilde {\mathcal G}}_{b}(\zeta)$, 
${\tilde {\mathcal G}}_{c}(\zeta)$, and 
${\tilde {\mathcal G}}_{d}(\zeta)$. These constraint functions have the vanishing Poisson brackets 
with the $2m - 4$ remaining ones ${\mathcal G}_{e}(\xi_{2}(\zeta))$ 
($e \ne a,b,c,d$). The latters constitute a 
complete non-degenerate set to determine the constraint submanifold ${\mathcal F}$ uniquely, and
so on. 

At the end, one gets in a neighborhood 
of $\xi$ a symplectic basis (\ref{TQ1}).

\subsection{Hamiltonian ${\tilde{\mathcal{H}}}$ constructed by gradient projection}

The Hamiltonian ${\tilde{\mathcal{H}}}$ of Sect. 4 can also be constructed with 
the help of Eq.(\ref{XO}): 
\begin{equation}
{\tilde{\mathcal{H}}}(\xi )=\mathcal{H}(\xi _{s}(\xi ))  \label{CANHAM}
\end{equation}
where $\xi _{s}(\xi )$ are gradient projections defined by ${\tilde{%
\mathcal{G}}}_{a}$.

Eq.(\ref{CANHAM}) and the algorithm described in Sect. 4 give, apparently, an equivalent 
Hamiltonian function, since the Hamiltonian flows on the constraint
submanifold coincide. It can be demonstrated by the comparison
of the Dirac brackets: 
\begin{equation}
\{ \xi^{i}, {\tilde {\mathcal{H}}}(\xi) \}_{D} = \{
\xi^{i}, \mathcal{H}(\xi) \}_{D} 
\end{equation}
which holds due to Eq.(\ref{DIBRXO}).

Applications of the gradient projection method to constructing the quantum 
deformation of the Dirac bracket can be found in \cite{QDDB}.

\section{Dirac quantization of spherical pendulum}
\setcounter{equation}{0} 

In Sect. 2, we modified the initial Lagrangian (\ref{LAGR}) on the
constraint submanifold $\chi = \ln\phi = 0$ to make more transparent the
origin of the underlying dilatation gauge symmetry. Let us formulate the Hamiltonian dynamics
of the spherical pendulum starting directly from Lagrangian 
$\mathcal{L} + \lambda \ln\phi$. The straightforward follow-up to the Dirac's
scheme (see also \cite{ABDA01,MFRF}) leads to the Hamiltonian dynamics described in Sect. 3:

Using the Dirac's scheme, we obtain the primary Hamiltonian $\mathcal{H}_{p} = 
\frac{1}{2}\pi^2 - \lambda \ln \phi$ and the primary constraint $\mathcal{G}%
_{0} = \pi_{\lambda} \approx 0$, where $\pi^{\alpha}$ are canonical momenta
associated to the canonical coordinates ${\phi}^{\alpha}$ and $%
\pi_{\lambda}$ is the canonical momentum associated to the Lagrange
multiplier $\lambda$. The canonical Hamiltonian becomes $\mathcal{H}_{c} = 
\mathcal{H}_{p} + u\pi_{\lambda}$ where $u$ is an undetermined function of
time. The secondary constraints $\mathcal{G}_{a+1} = \{ \mathcal{G}_{a}, 
\mathcal{H}_{c}\}$ can be found: $\mathcal{G}_{1} = \ln \phi$, $\mathcal{G}%
_{2} = \phi \pi/\phi^2$, $\mathcal{G}_{3} = \pi^2/\phi^2 - 2(\phi\pi)^2/
\phi^4 + \lambda/\phi^2$. The last constraint $\mathcal{G}_{3} = 0$ allows
to fix $\lambda$ in terms of $\phi^{\alpha}$ and $\pi^{\alpha}$, no new
constraints then appear.

The dimension of the phase space can be reduced by eliminating the
canonically conjugate pair $({\lambda},\pi_{\lambda})$. 
We solve equations $\mathcal{G}_{0} = 0$ with respect to $\pi_{\lambda}$ and $\mathcal{G}%
_{3} = 0$ with respect to $\lambda$ and substitute solutions into $%
\mathcal{H}_{c}$. The result is 
\begin{equation}
\mathcal{H}_{c}^{\prime} = \frac{1}{2}\pi^2 + (\pi^2 - 2(\phi\pi)^2/ \phi^2) \ln \phi. \label{HCP}
\end{equation}
 There remain two constraint
functions $\mathcal{G}_{1}$ and $\mathcal{G}_{2}$, such that $\{ \mathcal{G}%
_{1} ,\mathcal{G}_{2}\} = 1/\phi^2$. These are the second-class constraints.
The Hamiltonian $\mathcal{H}_{c}^{\prime}$ is first-class: $\{ \mathcal{G}%
_{1} ,\mathcal{H}_{c}^{\prime} \} = \mathcal{G}_{2} - \mathcal{G}_{1}\{ 
\mathcal{G}_{1} ,\lambda \}$ and $\{ \mathcal{G}_{2} ,\mathcal{H}%
_{c}^{\prime} \} = - \mathcal{G}_{1}\{ \mathcal{G}_{2} ,\lambda \}$ where $%
\lambda$ is determined by $\mathcal{G}_{3} = 0$.

In order to split the constraints into the gauge-fixing condition and the
gauge generator, we should construct, according to the previous Section, ${%
\tilde {\mathcal{G}}}_{a}$ and ${\tilde {\mathcal{H}}}$ starting from
$\mathcal{G}_{a}$ and $\mathcal{H}_{c}^{\prime}$.

In Sect. 3, we already had the same constraint submanifold described by
functions $\chi = \ln \phi$ and $\Omega = \phi \pi$ satisfying $\{ \chi
,\Omega \} = 1$, so one can set ${\tilde {\mathcal{G}}}_{1} = \ln \phi$ and $%
{\tilde {\mathcal{G}}}_{2} = \phi \pi$.

The Hamiltonian ${\tilde{\mathcal{H}}}$ can be constructed with the help of
Eq.(\ref{CANHAM}). Let us combine Eqs.(\ref{NEW3}), (\ref{NEW4}) and (%
\ref{NEW5}), (\ref{NEW6}) to get the full gradient projection: 
\begin{eqnarray}
\phi _{s}^{\alpha } &=&\phi ^{\alpha }/\phi , \\
\pi _{s}^{\alpha } &=&\phi \pi ^{\alpha }-\phi ^{\alpha }\phi \pi /\phi .
\end{eqnarray}
Replacing $\phi ^{\alpha }$ and $\pi ^{\alpha }$ in $\mathcal{H}_{c}^{\prime }$ 
by variables $\phi ^{\alpha }_{s}$ and $\pi ^{\alpha }_{s}$, respectively, we get 
\begin{equation}
{\tilde{\mathcal{H}}}=\frac{1}{2}\pi _{s}^{2} \label{HS}
\end{equation} 
which coincides with the Hamiltonian (\ref{HAM3}). 

{\it The difference ${\tilde{\mathcal{H}}}-\mathcal{H}_{c}^{\prime }$
is of the second order in the constraint functions. This is sufficient to
have identical Hamiltonian flows on the constraint submanifold.}

The equivalence is thus demonstrated, so further discussion can go in the
parallel with Sect. 3. In the example considered, the constraint
functions are of the second class, whereas in Sect. 3 they appear as the
gauge-fixing condition, $\chi = 0$, and the gauge generator, $\Omega$. This suggests
that the interpretation of second-class constraints of a holonomic system is a
matter of convention.

\section{Second-class constraints as generators of gauge symmetries
and gauge-fixing conditions}
\setcounter{equation}{0} 

The type of constraints appearing within the Hamiltonian framework depends on
the form of the corresponding Lagrangian. In case of the spherical pendulum,
starting from Eq.(\ref{LAGR}) one arrives to the constraints $\chi = 0$ and $%
\Omega = 0$ as to the second-class constraints of the Hamiltonian framework.
Starting from Eq.(\ref{LAG2}), the constraint $\chi = 0$ appears as a
gauge-fixing condition, whereas the constraint $\Omega = 0$ as a gauge
generator of the symmetry group. Lagrangians (\ref{LAGR}) and (\ref{LAG2})
are equivalent at the classical level, so they lead to the same
classical dynamics.

It is possible, therefore, at least in the case of spherical pendulum, 
to interpret second-class constraints as a gauge-fixing condition 
and a gauge generator, and vice versa. 

We wish to discuss whether this is a common situation. Any set of admissible 
gauge-fixing conditions and gauge generators can be treated as second-class 
constraints. This statement is widely used
for quantization of gauge theories (see, e.g., \cite{SLFA}). 

The reverse statement represents apparent interest also. The second-class 
constraints systems are analyzed from this point of view by Mitra and Rajaraman \cite{MITRA1}. 
In this Section, we discuss additional details about the underlying gauge symmetries of 
second-class constraints systems. 

\subsection{Converting a gauge system into a second-class constraints system}

Let a primary Hamiltonian $\mathcal{H}$ of a system be gauge invariant. The
generators of gauge transformations $\Omega_{A} = 0$ are such that 
\begin{eqnarray}
\{ \Omega_{A}, \mathcal{H} \} &=& \mathcal{R}_{A}^{\;B} \;\Omega_{B},
\label{1234} \\
\{ \Omega_{A}, \Omega_{B} \} &=& \mathcal{C}^{\;\;\;\;C}_{AB}\;\Omega_{C}.
\label{2345}
\end{eqnarray}
The gauge-fixing conditions $\chi_{A} = 0$ fulfill 
\begin{eqnarray}
\{ \chi_{A}, \chi_{B} \} &=& 0,  \label{3456} \\
\det \{ \chi_{A}, \Omega_{B} \} &\neq& 0.  \label{det}
\end{eqnarray}

After imposing gauge-fixing conditions, the system becomes equivalent to a
system described by a first-class Hamiltonian $\mathcal{H}^{\prime}$ defined
on the submanifold of second-class constraints $\mathcal{G}_{a} = (\chi_{A},
\Omega_{A})$. The Hamiltonian $\mathcal{H}^{\prime}$ can be constructed from
the canonical Hamiltonian of the original system, $\mathcal{H}_{c} = 
\mathcal{H} + \lambda _{B} \Omega_{B}$, by requiring 
\begin{equation}
\{ \chi_{A}, \mathcal{H}_{c} \} = \{\chi_{A}, \mathcal{H}\} + \lambda _{B}
\{ \chi_{A}, \Omega_{B}\} \approx 0.
\end{equation}
The gauge parameters $\lambda _{B}$ can be fixed from this equation in terms
of the canonical variables. We thus get the first-class Hamiltonian $%
\mathcal{H}^{\prime} = \mathcal{H}_{c}$ and second-class
constraints $\mathcal{G}_{a}$ satisfying $\{ \mathcal{G}_{a}, \mathcal{H}%
^{\prime}\} \approx 0$ and $\det \{ \mathcal{G}_{a}, \mathcal{G}_{b}\} \ne 0$.

The gauge-fixing is equivalent to converting the original
system into an equivalent one described by a first-class Hamiltonian and
second-class constraints.

\subsection{Converting a second-class constraints system into a gauge system}

It was shown by Dirac \cite{DIRAC1,DIRAC2} that first-class constraints
imply the presence of unphysical degrees of freedom the evolution of which 
is not fixed by the Hamilton's equations. The dynamics is self-consistent 
provided the unfixed degrees of freedom do not belong to the set of 
physical observables. This is the case of gauge-invariant systems which 
have gauge degrees of freedom. According to the Dirac's constraint dynamics, 
having first-class constraints is a necessary condition for the existence 
of a gauge symmetry. The reverse statement is less obvious:

Gauge transformations of the canonical coordinates have the form
\begin{equation}
\delta \phi ^{\alpha }=\{\phi ^{\alpha },\Omega _{A}\}\theta _{A} \label{GTRA}
\end{equation}
where $\theta _{A}$ are infinitesimal functions of time. Since $\Omega _{A}$
depend on the canonical coordinates and momenta, the variations $\delta \phi
^{\alpha }$ depend on the coordinates and momenta also, unless $\Omega _{A}$ 
are first degree polynomials of the canonical momenta. Subject to a Legendre
transform, $\delta \phi ^{\alpha }$ become, in general, functions of the
coordinates and velocities. The gauge theories such as QED and QCD have the 
constraint functions $\Omega _{A}$ as the first degree polynomials of
the canonical momenta. In such cases $\delta \phi ^{\alpha }=\delta \phi
^{\alpha }(\phi ^{\beta })$ do not depend on velocities and define thereby 
transformations on the configuration space ${\mathcal M} = (\phi^{\alpha})$ 
which induce, in turn, transformations in the tangent bundle 
$T{\mathcal M} = (\phi^{\alpha},{\dot \phi}^{\alpha})$.

The first-class constraints systems correspond to more general class of 
gauge-invariant systems with 
$\delta \phi ^{\alpha }=\delta \phi ^{\alpha }(\phi ^{\beta },\dot{\phi}^{\beta })$
and, probably, new auxiliary variables. 
In what follows, we distinguish thereby between gauge transformations 
of the form $\delta \phi ^{\alpha }=\delta \phi ^{\alpha }(\phi^{\beta })$ and 
generalized gauge transformations with velocity dependent parameters and/or 
new auxiliary variables. 

\textit{Having a first-class constraint system is a necessary, but not a
sufficient condition for an equivalent gauge-invariant system to exist
in the original configuration space. }

The possibility of constructing gauge-invariant systems in the unconstrained 
phase space, equivalent to second-class constraints systems upon a gauge-fixing, 
has been analyzed by Mitra and Rajaraman \cite{MITRA1}. 

The applications discussed \cite{MITRA1,VYTHE1,VYTHE2} have constraint 
functions, associated to gauge transformations, as polynomials of 
the canonical momenta of the degree less or equal to unity. The gauge transformations 
are not velocity dependent, although involve new auxiliary variables.
The gauged systems require generally an extended configuration space.

{\it If global symplectic basis in the space of constraint functions can 
be found, equivalent generalized gauge systems for second-class constraints systems
can be constructed in the unconstrained phase space.}

By passing over to the standard canonical frame, one can always select
${\tilde {\Omega}}_{A}$ as first canonical momenta. This is, however, not 
sufficient to have gauge invariance on the physical configuration space. 
Before doing a Legendre transform, one has to pass first to 
a canonical coordinate system where the coordinates $\phi^{\alpha}$ constitute a physical configuration 
space. This will mix up the canonical coordinates and momenta, preventing
from having ${\tilde {\Omega}}_{A}$ as the first degree polynomials.

It is worthwhile to notice that gauge invariant observables are not 
measurable if they involve auxiliary degrees of freedom. The sums of the 
vector potentials of the massive electrodynamics and the derivatives of the 
St\"uckelberg scalar are gauge invariant. They do not belong, however, 
to the set of physical observables. The equivalence with the ordinary 
gauge systems, where sets of gauge invariant quantities and physical observables 
coincide, is therefore not complete.

\subsection{Gauge-invariant systems as holonomic systems}

The quantization of gauge theories which appear under first-class constraints is studied
in many details (see, e.g., Refs. \cite{HENN,SLFA}). The second-class constraints systems are
usually quantized by employing the BFT formalism \cite{HENN,BATA,BATY}
that converts second-class constraints into a first class by extending the
original phase space to a higher dimension.

The BFT algorithm combined with the Fradkin-Vilkovisky quantization scheme \cite{FRVI,FRAD}
if applied to a system in a $2n$-dimensional phase space
ends up with an extended phase space of a dimension larger or equal to $2n + 12m$, since
the $2m$ constraints convert into first-class
constraints, each of which requires an independent gauge-fixing condition.
The remaining at least $8m$ degrees of freedom appear as ghost variables.

The problem of constructing gauge-invariant systems equivalent to
second-class constraints systems in the original phase space is 
discussed in Ref. \cite{MITRA1}. The existence of 
the global symplectic basis for the constraint functions is specified as 
sufficient conditions for the existence of gauged partners associated to the second-class 
constraints systems. 
The method \cite{MITRA1}, being successful in removing auxiliary fields 
from the original phase space, generates auxiliary
fields in the effective Lagrangians.

The systems of pointlike particles under
holonomic constraints have the natural gauge counterparts both
in the phase space and configuration space \cite{MFRF}.

In holonomic systems, second-class constraints split into
gauge-fixing conditions, ${\tilde \chi}_{A} = 0$, and gauge generators, ${\tilde
\Omega}_{A}$. Such systems can be quantized further as gauge theories. The
BFT method, if applied, would result to an extended phase space of the
dimension $2n + 6m$, containing $2m$ Lagrange multipliers and $4m$ ghost
variables. The underlying gauge symmetry allows to reduce the number of auxiliary fields 
within the quantization scheme \cite{FRVI,FRAD}.

\subsection{Frobenius' condition for holonomic constraints}

The holonomic constraints are defined by a constraint submanifold 
${\mathcal N} \subset {\mathcal M}$ in the configuration space ${\mathcal M}$. It comes 
then out automatically that particle velocities belong to the tangent space 
$T{\mathcal N}$ of the constraint submanifold.

An  $n - m$-dimensional tangent subspace $T{\mathcal N}  \subset T{\mathcal M}$ of an 
$n$-dimensional configuration space ${\mathcal M}$ can be defined in general through $m$ covector fields 
$\omega_{A \alpha}$ with $A=1,...,m$ and 
$\alpha =1,...,n$ by imposing constraints
\begin{equation}
\omega_{A \alpha}{\dot \phi}^{\alpha} = 0 \label{TANGENT}
\end{equation}
which are velocity dependent and therefore non-holonomic.

The Frobenius' integrability conditions (see, e.g., \cite{ARNO}),
\begin{equation}
\frac{\partial \omega_{A \alpha}}{\partial \phi^{\beta}}
({\dot \phi}_1^{\alpha}{\dot \phi}_2^{\beta} - {\dot \phi}_2^{\alpha}{\dot \phi}_1^{\beta})=0, \label{FROB}
\end{equation} 
for tangent vectors ${\dot \phi}_1^{\alpha}$ and ${\dot \phi}_2^{\beta}$ satisfying (\ref{TANGENT}), 
if fulfilled, implies the existence of a submanifold 
${\mathcal N}$ the tangent space of which coincides
with the tangent subspace (\ref{TANGENT}). In such a case, the constraints 
(\ref{TANGENT}) can be replaced with holonomic
constraints ${\chi}_{A} = 0$, identifying the constraint submanifold
${\mathcal N} \subset {\mathcal M}$, without modifying the dynamics. 

Eqs.(\ref{FROB}) specify non-holonomic systems which have gauged 
counterparts in the original configuration space. 

For an $n=3$ Euclidean space, Eqs.(\ref{TANGENT}) define, for $m=1$, a submanifold 
orthogonal to the vector $\omega_{\alpha}$. Eqs.(\ref{FROB}) 
tell us that the vector field ${\vec \omega}$ is a potential field provided that 
$rot{\vec \omega} = 0$. One can find a potential function ${\chi}$ such that 
${\vec \omega} = {\vec \nabla}\chi$. The sets of 
${\chi}= {\tt constant}$ define, in our case, possible constraint submanifolds of an equivalent holonomic system.

The tangent subspace determined by Eq.(\ref{CON1}) coincides with 
the tangent space of $S^{n-1}$. 
The covector field $\omega_{\alpha} = \phi^{\alpha}$ satisfies the Frobenius' condition.

\section{Supplementary conditions for Wigner functions}
\setcounter{equation}{0} 

The splitting of second-class constraints into constraints associated to 
gauge transformations and gauge-fixing conditions is not unique. Every pair 
($\chi_{A}$,$\Omega_{A}$) can be transformed, e.g., as $\chi_{A} \rightarrow
\Omega_{A}$, $\Omega_{A} \rightarrow - \chi_{A}$. The symplectic
group $Sp(2m)$ of transformations mixes $\chi_{A}$ and $\Omega_{A}$ further,
keeping the Poisson brackets invariant. Canonical transformations, furthermore,
do not modify the Poisson bracket of the constraints also. There exists therefore an 
affluent spectrum of representations and we have to demonstrate that the physical 
content of theory is invariant with respect to the choice of representation.

In the general case of second-class constraints, we do not have any
criterion how to discriminate between first-class
constraints associated to gauge transformations and gauge-fixing conditions. 
We show, conversely, that the way the second-class constraints are split
is not important for quantization.

A useful hint towards that conclusion is delivered by the path integral representation 
for the evolution operator (\ref{PARQ2}) which is invariant with respect to linear transformations
of  ${\mathcal G}_{a}$. 

The problem is to demonstrate that the supplementary conditions ${\hat \Omega}_{A} \Psi = 0$ 
can be replaced by ${\hat \chi}_{A} \Psi = 0$. In case of the
spherical pendulum, this is almost evident, since ${\hat \Omega} \Psi =
0$ means $\Psi \sim \Psi(\phi^{\alpha}/\phi)$ whereas ${\hat \chi} \Psi = 0$ means $%
\Psi \sim \delta(\chi) \Psi(\phi^{\alpha}) = \delta(\chi)
\Psi(\phi^{\alpha}/\phi)$. The essential dependence comes from the angular
part of the wave function, while the radial part $\delta(\chi)$ is
factorized, being commutative, Eq.(\ref{XIHA}), with the Hamiltonian and the $%
S $-matrix. It can be absorbed thereby by an overall normalization factor.

These arguments can apparently be extended to an arbitrary case.
According to the Dirac's prescription, the physical wave functions are
annihilated by the constraint operators associated to gauge transformations
\begin{equation}
{\hat \Omega}_{A}\Psi = 0.  \label{DIRO}
\end{equation}

To simplify the matter, we pass to the standard canonical coordinate system and 
choose the constraint functions $\chi_{A}$ as first $m$ canonical coordinates 
$q_{A}$ and the constraint functions $\Omega_{A}$ as first $m$ canonical momenta 
$p_{A}$. The remaining canonical variables 
are ($q^{*},p^{*}$) constitute the physical phase space $\Gamma^{2(n-m)}_{*}$. 
Eq.(\ref{DIRO}) gives then, in the coordinate representation, a wave function of the form 
$\Psi = \Psi(q^{*})$. Such a wave function has an infinite norm, since the
integral $\int d^{n}q|\Psi(q^{*})|^2$ diverges. In the momentum representation, 
we would get $\Psi(p) = (2\pi \hbar)^{m} \prod_{A} \delta(p_{A}) \Psi(p^{*})$. $\Psi(p)$ has an
infinite norm $(2\pi \hbar)^{m} \delta^{m}(0)$ provided $\Psi(p^{*})$ is normalized to unity. 
An infinite but numerical factor can be included into the norm of $\Psi(p)$.

Let us check whether wave functions satisfying 
\begin{equation}
{\hat \chi}_{A}\Psi = 0.  \label{DIRX}
\end{equation}
have a physical sense. In the coordinate representation, one gets $\Psi = \prod_{A} \delta(q_{A})
\Psi(q^{*}) $ and in the momentum space $\Psi = \Psi(p^{*})$. These states
have infinite norms. There is an apparent symmetry $p \rightarrow q$, $q
\rightarrow - p$ between wave functions satisfying (\ref{DIRO}) and (\ref
{DIRX}). 

Owing to the factor $\prod_{A} \delta(q_{A})$, conditions (\ref
{DIRO}) and (\ref{DIRX}) single out the same set of functions, the nontrivial
dependence of which comes from the physical variables $q^{*}$ ($p^{*}$) only.

The Wigner functions \cite{WIGNER} provide complete information on quantum systems. 
The association rules of operators in the Hilbert space and functions 
in the phase space are discussed a long time \cite{WEYL,WIGNER,GROE,MOYAL}. The quantum mechanics 
can be reformulated using the Groenewold star-product \cite{GROE} representing 
a deformation of the usual pointwise product of functions in the phase space. The Wigner functions 
and constraint functions are defined in the phase space, so it is natural to 
discuss supplementary conditions in terms of the Wigner functions. In this approach, 
the equivalence between Eqs.(\ref{DIRO}) and (\ref{DIRX}) becomes more transparent. 
In addition, more general group of supplementary conditions, equivalent to Eqs.(\ref{DIRO}) 
and (\ref{DIRX}), can be formulated in terms of the Wigner functions.

\subsection{Probability density localized on the constraint submanifold}

Let us start from the standard canonical frame \cite{MASKAWA} where the constraint functions 
$\chi_{A}$ are the first $m$ canonical coordinates $q_{A}$ and the constraint 
functions $\Omega_{A}$ are the first $m$ canonical momenta $p_{A}$. The 
probability density in the phase space $\Gamma^{2n}$ can be written as follows: 
\begin{equation}
{W}(q,p) = (2\pi \hbar)^{m} \prod_{A} \delta(q_{A}) \delta(p_{A}) W_{*}(q^{*},p^{*}).  \label{DEDM}
\end{equation}
Identifying the $W(q,p)$ with the Wigner function in the unconstrained phase space 
and using the Wigner transform, one gets the density matrix 
\begin{eqnarray}
{\rho}(q,q^{\prime}) &=& \int W(\frac{q + q^{\prime}}{2},p) e^{\frac{i}{\hbar}p(q -
q^{\prime})} \frac{d^n p}{(2\pi\hbar)^{n}}  \nonumber \\
&=& \prod_{A} \delta(\frac{q_{A} + q^{\prime}_{A}}{2}) \rho_{*}(q^{*},q^{*
\prime}).  \label{DENM}
\end{eqnarray}
It satisfies the operator equations 
\begin{eqnarray}
{\hat q}_{A} {\hat{\rho}} + {\hat{\rho}} {\hat q}_{A} = 0, \label{ONS1} \\
{\hat p}_{A} {\hat{\rho}} + {\hat{\rho}} {\hat p}_{A} = 0. \label{ONS2}
\end{eqnarray}
The first equation implies $\rho \sim \delta(q_{A} + q^{\prime}_{A})$. 
The second one implies that $\rho$ does not depend on $q_{A} - q^{\prime}_{A}$. 

The density matrix $\rho_{*}(q^{*},q^{* \prime})$ is normalized to unity, so 
\begin{equation}
\int{\ \rho(q,q) d^n q} = 1. \label{NC1}
\end{equation}

One can make a unitary transformation to pass to an arbitrary set of 
operators associated to the canonical variables. In terms of 
\[
{\hat {\mathcal G}}_{a} = 
{\mathcal U}({\hat q}_{A},{\hat p}_{A}){\mathcal U}^{+},
\]
and ${\hat \rho}$ replaced with  ${\mathcal U} {\hat \rho} {\mathcal U}^{+}$,
Eqs.(\ref{ONS1}) and (\ref{ONS2}) become 
\begin{equation}
{\hat {\mathcal G}}_{a} {\hat \rho} + {\hat \rho} {\hat {\mathcal G}}_{a} = 0.  \label{ANTI}
\end{equation}

Eqs.(\ref{ANTI}) are necessary and sufficient conditions to have the 
representation (\ref{DENM}) in the standard canonical frame. 

{\it The supplementary conditions (\ref{ANTI}) cannot be formulated in terms of a wave function,
since the density matrix in the unconstrained configuration space does not correspond to a pure 
state, even if system on the constraint submanifold is in a pure state.}

In an arbitrary frame and in the classical limit, Eq.(\ref{DEDM}) looks like 
\begin{equation}
W(\xi) = (2\pi\hbar)^{m} \prod_{a} \delta(\mathcal{G}_{a}) \sqrt{\det \{ 
\mathcal{G}_{a}, \mathcal{G}_{b} \}} W_{*}(\xi_{s}(\xi)). \label{DEDMOD}
\end{equation}
Note that $\prod_{a} \delta(\mathcal{G}_{a})$ 
acts as a projection operator, so that $
\prod_{a} \delta(\mathcal{G}_{a}) f(\xi) = \prod_{a} \delta(\mathcal{G}_{a})
f(\xi_{s}(\xi))$ $\forall f(\xi)$. 

In the classical limit, the Wigner function satisfies
\begin{equation}
{\mathcal{G}_{a}}(\xi)W(\xi) = 0. \label{DE3}
\end{equation}

The complete phase-space analogue of quantum Eqs.(\ref{ANTI}) can be formulated 
in terms of the symmetric part of the Groenewold star-product \cite{GROE} to give
\begin{equation}
{\mathcal{G}_{a}}(\xi)\circ W(\xi) = 0. \label{DE4}
\end{equation}
The star-product has the following decomposition:
\begin{equation}
f(\xi)\star g(\xi)=f(\xi)\circ g(\xi)+\frac{i\hbar }{2}f(\xi)\wedge g(\xi)  \label{SSM}
\end{equation}
where
\begin{eqnarray}
f(\xi )\circ g(\xi ) &=&f(\xi )\cos ( \frac{\hbar }{2}\mathcal{P}) g(\xi ),  \label{ES} \\
f(\xi )\wedge g(\xi ) &=&f(\xi )\frac{2}{\hbar }\sin ( \frac{\hbar }{2}%
\mathcal{P}) g(\xi )  \label{EA}
\end{eqnarray}
and
\[
\mathcal{P} = -{I}^{ij} \overleftarrow{\frac{%
\partial }{\partial \xi ^{i}}}\overrightarrow{\frac{\partial }{\partial \xi
^{j}}} 
\]
is the so-called Poisson operator. In the limit $\hbar \rightarrow 0$, 
\begin{equation}
\lim_{\hbar \rightarrow 0}f(\xi )\wedge g(\xi )=\{f(\xi ),g(\xi )\}. \label{LIMI}
\end{equation}
The Poisson bracket $\{f(\xi ),g(\xi )\}$ coincides generally with a
function associated to the commutator $-i/\hbar [\hat{f},\hat{g}]$ to
the lowest order in the Planck's constant only. The skew-symmetric part 
of the Groenewold product provides a
generalization of the Poisson bracket which is skew-symmetric with respect to
two functions, real for real functions, coincides with the Poisson bracket
to the lowest order in the Planck's constant, satisfies the Jacoby
identity, and keeps the association rule for commutators. This part of the Groenewold product
is known also as the "sine bracket" or Moyal bracket \cite{MOYAL}. 

In the limit $\hbar \rightarrow 0$, the quantum condition (\ref{DE4})
recovers the classical one (\ref{DE3}). These two conditions coincide
in the standard canonical frame.

The normalization condition Eq.(\ref{NC1}) holds in the quantum case. The
Wigner function satisfies
\begin{equation}
\int{W(\xi) \frac{d^{2n}  \xi}{(2\pi \hbar)^{n}}} = 1. \label{NC2}
\end{equation}
The classical limit of the Wigner function Eq.(\ref{DEDMOD}) provides the 
conventional normalization on the constraint submanifold for $W_{*}(\xi_{s}(\xi))$.

The Groenewold star-product is non-local. The usual pointwise product of two functions
like in Eq.(\ref{DE3}) has no quantum counterpart. It can be treated as a classical 
limit of a quantum operator equation only. By contrary, Eq.(\ref{DE4}) is the 
phase space analogue of the quantum operator equation (\ref{ANTI}).

Eq.(\ref{DEDM}) accomplishes a trivial
extrapolation of the Wigner function from the constraint submanifold: The 
density is set equal to zero when $\xi$ does not belong to the 
constraint submanifold. 
The Wigner function is not a smooth function across the constraint 
submanifold, so it is hard to formulate using this approach an evolution equation 
similar to the Liouville equation in the unconstrained phase space.

\subsection{Probability density localized on and outside of the constraint submanifold}

One can require that at any given point $\xi $ the density be the same as at 
$\xi _{s}=\xi _{s}(\xi )$. The gradient projections $\xi _{s}$
can be constructed, as discussed in Sect. 5, using phase flows generated by
the constraint functions $\mathcal{G}_{a}$ to solve equations 
$\mathcal{G}_{a}(\xi _{s}(\xi ))=0$.

The analogue of Eq.(\ref{DEDM}) reads 
\begin{equation}
W(\xi) = W_{*}(\xi_{s}(\xi)). \label{DEDN}
\end{equation}
Such a density has infinite norm, since there are directions in the phase
space $\Gamma^{2n}$, crossing the submanifold $\mathcal{G}_{a} = 0$, along which
the density remains constant. These directions are determined by the constraint
functions. 

Eq.(\ref{XO}) tells 
\begin{equation}
\{ \mathcal{G}_{a}(\xi), W(\xi)\} = 0.  \label{1111}
\end{equation}

In the standard canonical frame, where $q_{A}=\chi _{A}$ and $p_{A}=\Omega _{A}$, the
gradient projections can be easily constructed: 
\begin{equation}
\xi _{s}(\xi)=(0,q^{*},0,p^{*})
\end{equation} 
where $\xi =(q_{A},q^{*},p_{A},p^{*})$. Eq.(\ref{DEDN}) then simplifies to 
\begin{equation}
W (q,p)=W_{*}(q^{*},p^{*}).  \label{DEDB}
\end{equation}
Note that \begin{eqnarray}
\xi _{u}(\xi )&=&(0,q^{*},p_{A},p^{*}), \\
\xi _{v}(\xi)&=&(q_{A},q^{*},0,p^{*}),
\end{eqnarray}
so that $\xi _{s}(\xi ) = \xi _{u}(\xi _{v}(\xi ))=\xi_{v}(\xi _{u}(\xi ))$.

If we apply the Wigner transform, we get the density matrix 
\begin{eqnarray}
{\rho}(q,q^{\prime}) &=& \int \rho(\frac{q + q^{\prime}}{2},p) e^{\frac{i}{\hbar}p(q -
q^{\prime})} \frac{d^n p}{(2\pi\hbar)^{n}}  \nonumber \\
&=& \prod_{A} \delta(q_{A} - q^{\prime}_{A}) \rho_{*}(q^{*},q^{* \prime}).
\label{DENB}
\end{eqnarray}
It satisfies 
\begin{eqnarray}
{\hat q}_{A} {\hat {\rho}} - {\hat {\rho}} {\hat q}_{A} &=& 0, \\
{\hat p}_{A} {\hat {\rho}} - {\hat {\rho}} {\hat p}_{A} &=& 0
\end{eqnarray}
or, equivalently,
\begin{equation}
{\hat {\mathcal G}}_{a} {\hat \rho} - {\hat \rho}{\hat  {\mathcal G}}_{a} = 0.  \label{COMM}
\end{equation}
The phase space analogue of these operator equations looks like
\begin{equation}
\mathcal{G}_{a}(\xi) \wedge W(\xi) = 0.  \label{COMM1}
\end{equation}
This condition is in agreement with its classical counterpart 
Eq.(\ref{1111}) by virtue of Eq.(\ref{LIMI}).

Eqs.(\ref{COMM}) are distinct from Eqs.(\ref{ANTI}). This is a consequence 
of different extrapolation schemes of the density to the unconstrained phase space. 
Eqs.(\ref{COMM}) are the necessary and sufficient conditions for Eq.(\ref{DENB}).

The dependence of the density matrix on ($q_{A},q_{A}^{\prime}$) does not factorize, 
although $\rho^2 = \rho$ for $\rho_{*}^2 = \rho_{*}$. The latter, even when fulfilled, 
is not sufficient to have a pure state. Eq.(\ref{DENB}) describes, in particular, a
noncoherent sum of pure states with different momenta $p_{A}$. The system is
thereby identified as a mixed state. It cannot be described by a wave
function. The constraint equations (\ref{COMM}), respectively, cannot be formulated in
terms of a wave function $\Psi(q)$.

If one works with density matrices in the representation (\ref{DEDM}) or (%
\ref{DEDB}), it is not important how the second-class constraints were split.
Eqs.(\ref{ANTI}) and (\ref{COMM}) are symmetric explicitly with respect to
the interchange $\chi_{A} \rightarrow \Omega_{A}$, $\Omega_{A} \rightarrow -
\chi_{A}$, linear transformations of the 
constraint functions, and furthermore, with respect to unitary transformations in the Hilbert space.
In the classical limit, they are invariant with respect to canonical transformations.

The normalization condition of the Wigner function involves a projection operator
\begin{equation}
\mathfrak{P} = \int \frac{d^{2m}\lambda }{(2\pi \hbar )^{m}}\prod_{a=1}^{2m}\exp (\frac{i}{\hbar }
\hat{\mathcal{G}}^{a}\lambda _{a})
\label{proj}
\end{equation}
in terms of which the norm is calculated as follows \cite{QDDB}
\begin{equation}
\int{P(\xi) \star W(\xi) \frac{d^{2n}  \xi}{(2\pi \hbar)^{n}}} = 1. \label{NC5}
\end{equation}
where $P(\xi)$ is the Weyl's symbol of the operator (\ref{proj}). Note that 
$\mathfrak{P}$ is commutative with $\hat{\rho}$ in virtue of (\ref{COMM}).

The Wigner function appears as a smooth function, so it is an appropriate 
object to describe an evolution of 
the system on line with the Liouville equation in the unconstrained phase 
space. A quantum extension of the 
the Liouville equation for $W(\xi)$ satisfying Eqs.(\ref{COMM1}) is given 
in Ref.\cite{QDDB}.

\subsection{Probability density with mixed localization}

Let us keep, e.g., canonical momenta on the
constraint submanifold and use the gradient projection to extend the density 
as a function of the canonical coordinates away from the constraint submanifold. 
The Wigner function is given then by 
\begin{equation}
W (q,p)=(2\pi \hbar)^{m}\prod_{A}\delta (p_{A})W_{*}(q^{*},p^{*}).
\label{2222}
\end{equation}
Applying the Wigner transform, we get the density matrix 
\begin{eqnarray}
{\rho}(q,q^{\prime}) = \rho_{*}(q^{*},q^{* \prime})  \label{DENC}
\end{eqnarray}
which satisfies 
\begin{eqnarray}
{\hat p}_{A} {\hat \rho} &=& 0, \\
{\hat \rho} {\hat p}_{A} &=& 0
\end{eqnarray} 
or, equivalently, 
\begin{eqnarray}
{\hat {\Omega}}_{A} {\hat \rho} &=& 0,  \label{COMC1} \\
{\hat \rho} {\hat {\Omega}}_{A} &=& 0.  \label{COMC}
\end{eqnarray}

These equations provide the necessary and sufficient conditions for Eq.(\ref
{DENC}). No dependence on the ($q_{A}$,$q_{A}^{\prime}$) appears. If $%
\rho_{*}^2 = \rho_{*}$ in the coordinate space $q^{*}$ on the constraint submanifold 
then $\rho^2 = \rho$ in the coordinate space $q$ (owing to an infinite normalization factor). 

{\it The system is allowed to
appear in a pure state and can be described by a wave function $\Psi(q)$ accordingly.
Eqs.(\ref{COMC1}) and (\ref{COMC}) can be reformulated in terms of
wave functions to match the Dirac's supplementary condition
Eq.(\ref{DIRO}).}

The mixed localization scheme breaks the symmetry 
$\chi_{A} \rightarrow \Omega_{A}$, $\Omega_{A} \rightarrow - \chi_{A}$ from
the outset. However, it allows to work with wave functions.
The other methods we discussed lead, in the unconstrained 
configuration space, to mixed states.

The quantum analogue of Eqs.(\ref{COMC1}) and (\ref{COMC}) in the phase space
can be formulated in terms of the star-product:
\begin{eqnarray}
{\Omega}_{A}(\xi) \star W(\xi) &=& 0, \\
W(\xi) \star {\Omega}_{A}(\xi) &=& 0.
\end{eqnarray}

As a consequence of Eqs.(\ref{COMC1}) and (\ref{COMC}), we have 
$\{\Omega_{A} , W \} = 0$ in the classical limit. Eq.(\ref{XX}) tells then that $W(\xi) =
\rho(\xi_{u}(\xi))$. The second classical condition $\Omega_{A} W = 0$
results in the Wigner function $W(\xi) = \prod
\delta(\Omega_{A}(\xi)) \rho_{*}(\xi_{v}(\xi))$. Combining these two
equations, we obtain
\begin{equation}
W(\xi) = (2\pi\hbar)^m \prod \delta(\Omega_{A}(\xi)) W_{*}(\xi_{s}(\xi)),
\label{AAAA}
\end{equation}
which is in agreement with Eq.(\ref{2222}). We used here $\Omega_{A}(\xi_{u}(\xi)) =
\Omega_{A}(\xi)$ which is valid up to the second order in the constraint
functions $\chi_{A}$.

It is possible to establish a relationship with the results
of the previous subsection. Indeed, one can check that
\begin{equation}
{\hat \rho} = \mathfrak{P}_{\Omega} {\hat \rho}_{s},
\label{relationship}
\end{equation}
with ${\hat \rho}_{s}$ being the density matrix from the previous subsection,
satisfies Eqs.(\ref{COMC1}) and (\ref{COMC}). The operator $\mathfrak{P}_{\Omega}$
is defined by
\begin{equation}
\mathfrak{P}_{\Omega} = \int \frac{d^{m}\lambda }{(2\pi \hbar )^{m}}\prod_{A=1}^{m}\exp (\frac{i}{\hbar }
\hat{\Omega}_{A}\lambda _{A})
\label{projomega}
\end{equation}

In order to get the normalization condition for $W(\xi)$ one has to 
factorize ${\hat \rho}$ according to (\ref{relationship}), 
construct the Weyl's symbol of ${\hat \rho}_{s}$, and apply Eq.(\ref{NC5}).

Let us consider the opposite localization: 
\begin{equation}
W(q,p) = \prod_{A} \delta(q_{A}) W_{*}(q^{*},p^{*}).  \label{DEND}
\end{equation}
The first $m$ canonical coordinates are kept on the constraint
submanifold, whereas the first $m$ canonical momenta are projected.

The density matrix has the form 
\begin{eqnarray}
{\rho}(q,q^{\prime}) = \prod_{A} \delta(q_{A}) \delta(q_{A}^{\prime})
\rho_{*}(q^{*},q^{* \prime})  \label{DENE}
\end{eqnarray}
It satisfies ${\hat q}_{A} {\hat \rho} = {\hat \rho} {\hat q}_{A} = 0$ or, equivalently, 
\begin{eqnarray}
{\hat {\chi}}_{A} {\hat \rho} &=& 0,  \label{COME1} \\
{\hat \rho}{\hat  {\chi}}_{A} &=& 0.  \label{COME2}
\end{eqnarray}
These conditions are the necessary and sufficient conditions to have the 
density matrix of the form (\ref{DENE}).

The system is allowed to appear in a pure state and can be described by a
wave function. The constraints (\ref{COME1}) and (\ref{COME2}) can be
reformulated in terms of the wave functions to give Eq.(\ref{DIRX}).

As a consequence of Eqs.(\ref{COME1}) and (\ref{COME2}), one gets in the 
classical limit 
\begin{equation}
W(\xi) = \prod \delta(\chi_{A}(\xi)) W_{*}(\xi_{s}(\xi)),  \label{BBBB}
\end{equation}
in agreement with Eq.(\ref{DEND}). We used here the relation $%
\chi_{A}(\xi_{v}(\xi)) = \chi_{A}(\xi)$ which is valid up to the second
order in $\Omega_{A}$.

The quantum analogue of Eqs.(\ref{COME1}) and (\ref{COME2})
is given by
\begin{eqnarray}
{\chi}_{A}(\xi) \star W(\xi) &=& 0, \\
W(\xi) \star {\chi}_{A}(\xi) &=& 0.
\end{eqnarray}

There exists a relationship with the Wigner function 
of the previous subsection. One can check that
\begin{equation}
{\hat \rho} = \mathfrak{P}_{\chi} {\hat \rho}_{s}
\label{relationship2}
\end{equation}
satisfies Eqs.(\ref{COME1}) and (\ref{COME2}). The operator $\mathfrak{P}_{\chi}$
is defined by
\begin{equation}
\mathfrak{P}_{\chi} = \int d^{m}\lambda \prod_{A=1}^{m}\exp (\frac{i}{\hbar }
\hat{\chi}_{A} \lambda_{A})
\label{projomega2}
\end{equation}
Note that $\mathfrak{P} = \mathfrak{P}_{\Omega}\mathfrak{P}_{\chi}$ and 
$\hat{\Omega}_{A}\mathfrak{P}_{\Omega} = \hat{\chi}_{A}\mathfrak{P}_{\chi} = 0$ 
in the operator sense.

The normalization of the Wigner function in arbitrary canonical 
coordinate system is quite involved:

{\it Given a wave function in the unconstrained configuration space, one should
construct the density matrix, factorize it according to 
Eqs.(\ref{relationship}) or (\ref{relationship2}) and integrate the
Wigner function associated to $\rho_{s}$ in order to extract the 
norm according to Eq.(\ref{NC5}).}

Respectively, the Hermitian product of wave functions is calculated using
the off-diagonal Wigner function 
and integrating it over the unconstrained phase space according to the
same prescription.

\subsection{Discussion}

In the classical constraint systems, physical quantities depend on the probability 
densities localized on the constraint submanifold only. In the standard canonical 
frame, the Wigner functions of quantum systems are localized on the constraint 
submanifolds also. This is mandatory, since any unconstrained system can be treated 
as a constrained system with constraints imposed to remove the added unphysical 
degrees of freedom. In doing so, the unphysical degrees of freedom should not modify 
dynamics of the initial system. This is achieved by attributing physical sense to the 
Wigner functions on the constraint submanifold. How to extrapolate Wigner functions 
from the constraint submanifold into the unconstrained phase space is a matter of 
convention.

We discussed the most evident extrapolations. Among them are
those which allow to describe systems in the original phase space as pure states
(mixed localization). One of them can be constructed using the Dirac's
prescription (\ref{DIRO}). The dual condition (\ref{DIRX}) was found to be
possible also. If we work with density matrices or the Wigner functions, one arrives 
at conditions (%
\ref{ANTI}) or (\ref{1111}), both are symmetric under the permutations $%
\chi_{A} \rightarrow \Omega_{A}$, $\Omega_{A} \rightarrow - \chi_{A}$. 
The constraints imposed on the physical states
do not depend on the splitting $\mathcal{G}_{a} = (\chi_{A}, \Omega_{A})$.

\section{The $O(n)$ non-linear sigma model}
\setcounter{equation}{0} 

The $O(n)$ non-linear sigma model represents the field theory analogue
of the spherical $n-1$-dimensional pendulum. The $n=4$ case corresponds to
the chiral non-linear sigma model due to the isomorphism of algebras 
$su(2)_{L}\oplus su(2)_{R}\sim so(4)$. 

In Sect. 2, we started from the tangent bundle
$T{\mathcal M} = (\phi^{\alpha},{\dot \phi}^{\alpha})$ of the dimension $2n$, 
defined over the configuration space ${\mathcal M} = (\phi^{\alpha})$. 
Lagrangian (\ref{LAGR}) depends on $n$ velocities ${\dot \phi}^{\alpha}$. 
On the constraint submanifold (\ref{CONS}), it depends on $n-1$ velocities 
$\Delta^{\alpha \beta}{\dot \phi}^{\beta}$ tangent to the constraint submanifold.
The dynamics of $\phi$ turns out to be independent on other variables. 
The constraint (\ref{CONS}) reduces the effective number of the
degrees of freedom to $n - 1$. Thus, a $2n - 2$ dimensional tangent bundle over the 
$n-1$ configuration space can be
constructed. It is described, e.g., by coordinates $\vartheta^i$ analogous 
to the angular coordinates in three dimensional space. Lagrangian (\ref
{LAG2}) is not degenerate with respect to ${\dot \vartheta}^i$. The
coordinates $\vartheta^i$ constitute the physical configuration space. 

The path integral for the evolution operator in terms of the 
angular coordinates $\vartheta^i$ can be treated as a reference point for
comparison to more involved quantization methods.


\subsection{Path integral for the $O(n)$ non-linear sigma model}


Let us construct the path integral for the $O(n)$ non-linear sigma model in terms 
of the angular variables $\vartheta^i$, i.e., by solving the constraint equations from the 
outset and compare it with the expression of Sect. 2 derived using
the underlying gauge symmetry of the $O(n)$ non-linear sigma model.

The field theory extension of the spherical pendulum problem is rather straightforward. 
In what follows, the kinematic variables are functions of $x_{\mu} = (t,\mathbf{x})$.

Let us construct a basis 
\begin{eqnarray}
e^{\alpha}_i &=& \frac{\partial } {\partial \vartheta^i} \phi^{\alpha},
\label{E1} \\
e^{\alpha}_i e^{\alpha}_j &=& g_{ij},  \label{E2} \\
g^{ij} e^{\alpha}_i e^{\beta}_j &=& \delta^{\alpha \beta} - \phi^{\alpha}
\phi^{\beta}/\phi^2,  \label{E3}
\end{eqnarray}
where $g_{ij} = g_{ij}(\vartheta)$ is an induced metric tensor on the submanifold $%
\phi(x) = 1$, $\det || g_{ij} ||\neq 0$. In terms of the coordinates $%
\vartheta^i$, the field theory extension of Lagrangian (\ref{LAG2}) takes
the form 
\begin{equation}
\mathcal{L_{*}} = \frac{1}{2} g_{ij}{\partial_{\mu} \vartheta ^i}{%
\partial_{\mu} \vartheta ^j}.  \label{LAG4}
\end{equation}

The Legendre transformation of $\mathcal{L_{*}}$ is well defined. In
term of the canonical momenta 
\begin{equation}
\varrho _{i}=\frac{\partial \mathcal{L_{*}}}{\partial {\dot{\vartheta}}^{i}}%
=g_{ij}{\dot{\vartheta}}^{j}  \label{MOME}
\end{equation}
the Hamiltonian density can be found to be 
\begin{equation}
\mathcal{H_{*}}=\frac{1}{2}g^{ij}\varrho _{i}\varrho _{j}+\frac{1}{2}g_{ij}{%
\partial _{a}\vartheta ^{i}}{\partial _{a}\vartheta ^{j}}  \label{HAMI}
\end{equation}
where $a=1,2,3$. The non-vanishing Poisson bracket for the canonical
coordinates and momenta has the form 
\begin{equation}
\{\vartheta ^{i}(t,\mathbf{x}),\varrho _{j}(t,\mathbf{x}^{\prime })\}=\delta
_{j}^{i}\delta (\mathbf{x}-\mathbf{x}^{\prime }).  \label{POIS}
\end{equation}
The Poisson bracket relations for coordinates $\phi ^{\alpha }$ and momenta $\pi ^{\alpha }$
associated to the tangent velocities $\pi_{\perp} ^{\alpha }=\varrho
_{i}g^{ij}e_{j}^{\alpha }$ agree with those discussed in Sect. 4.

The path integral in the phase space $(\vartheta^i, \varrho_i)$ is given by
\begin{equation}
Z = \int{\ \prod _{} \frac{d\vartheta^i d\varrho_i}{(2\pi\hbar)^{n-1}} \exp \left \{ \frac{i}{\hbar} \int{%
d^{4}x(\varrho^i {\dot \vartheta_i} - \mathcal{H_{*}})} \right \} }.
\label{PART}
\end{equation}
The Liouville measure $d\vartheta^i d\varrho_i$ is consistent with Eq.(\ref{POIS}).
The integral over the canonical momenta in Eq.(%
\ref{PART}) has a Gaussian form and can be calculated
explicitly: 
\begin{equation}
Z = \int{\ \prod _{} \sqrt{\det || g_{ij} ||} d^{n-1}\vartheta \exp \left \{
\frac{i}{\hbar} \int{dx^{4}\mathcal{L_{*}}} \right \} }.  \label{MEAS}
\end{equation}
The value $\sqrt{\det || g_{ij} ||} d^{n - 1}\vartheta$ gives
volume of the configuration space, defined by the metric tensor $g_{ij}$. This
measure is invariant under the $O(n)$ group.

The $S$-matrix (\ref{MEAS}) can be written in an explicitly covariant form
with respect to the $O(n)$ rotations and the Lorentz transformations. First,
we rewrite Lagrangian density (\ref{LAG4}) in terms of the coordinates $%
\phi^{\alpha}$ 
\begin{equation}
\mathcal{L}_{*} = \frac{1}{2} \Delta^{\alpha \beta}(\phi){{\partial_{\mu}}{%
\phi}^{\alpha} {\partial_{\mu}}{\phi}^{\beta}}/{\phi^{2}}  \label{LAG7}
\end{equation}
and, second, rewrite the Lagrange measure
\begin{equation}
\sqrt{\det || g_{ij} ||} d^{n-1}\vartheta = \sqrt{\left ({\partial \chi}/{%
\partial \phi^{\alpha}} \right )^2}\delta(\chi)d^n\phi
\end{equation}
where $\chi = \ln\phi$. The right side is the same for all functions
vanishing at $\phi = 1$.

In this form we recover the result of Sect. 2.


\subsection{Pion field parameterization in chiral sigma model}


The $n=4$ case is especially interesting since it corresponds to the chiral
non-linear sigma model. For $n=4$, the angular coordinates are
defined by 
\begin{equation}
\phi^{\alpha} = (\cos\psi,\sin\psi \times (\cos\theta, \sin\theta \times
(\cos\varphi, \sin\varphi)))
\end{equation}
where $\vartheta^1 = \psi$, $\vartheta^2 = \theta$, and $\vartheta^3 =
\varphi$.

The angular distance, $\Theta$, between two vectors $\phi^{\alpha}$ and $%
\phi^{\prime \alpha}$ is defined by scalar product $\cos\Theta = \phi
\phi^{\prime}$. The distance element becomes 
\begin{equation}
d\Theta^2 = d\psi^2 + \sin^2\psi (d\theta^2 + \sin^2\theta d\varphi^2).
\end{equation}
The components of the metric $g_{ij}$ can be found using the expansion $%
d\Theta^2 = g_{ij} d\vartheta^{i} d\vartheta^{j}$ or directly from Eqs.(\ref
{E1}) and (\ref{E2}). The Lagrange measure of the path integral becomes 
\begin{eqnarray}
\sqrt{\det || g_{ij} ||} d^3\vartheta &=& \sin^2\psi \sin\theta d\psi
d\theta d\varphi  \nonumber \\
&=& \frac{ \sin^2\psi }{ \psi^2 }dV,
\end{eqnarray}
with $dV = \psi^2 d\psi \sin\theta d\theta d\varphi$ being an element of the
Euclidean volume. The corresponding Lagrangian can be found from Eq.(%
\ref{LAG4}) to give
\begin{equation}
\mathcal{L}_{*} = \frac{1}{2}(\partial_{\mu}\psi)^2 + 
\frac{\sin^2\psi}{2} ((\partial_{\mu} \theta)^2 + \sin^2\theta
(\partial_{\mu} \varphi)^2).
\end{equation}

The quantization of the chiral sigma model is made using an oscillator basis
by expanding the $\mathcal{L}_{*}$ around $\vartheta^{i} = 0$ breaking
thereby the $O(4)$ symmetry down to its $O(3)$ subgroup. It can be
successful provided that measure of the coordinate space is such that $\det
|| g_{ij} || = 1$. The path integrals convert then to the Gaussian integrals
which can be calculated. The parameterization preserving the $O(3)$ 
symmetry and satisfying the above requirement is,
apparently, unique. One should rescale the "radius" $\psi$ according to
\begin{equation}
\sin^2\psi d\psi = \omega^2 d\omega.
\end{equation}
This elementary equation gives 
\begin{equation}
\omega = \left (\frac{3}{2}(\psi - \sin\psi \cos\psi) \right )^{1/3}.
\label{NPAR}
\end{equation}
Lagrangian $\mathcal{L}_{*}$ then becomes 
\begin{equation}
\mathcal{L}_{*} = \frac{\omega^4}{2\sin^4\psi}(\partial_{\mu}\omega)^2 + 
\frac{\sin^2\psi}{2} ((\partial_{\mu} \theta)^2 + \sin^2\theta
(\partial_{\mu} \varphi)^2)
\end{equation}
where $\psi$ is a function of $\omega$ Eq.(\ref{NPAR}). The mass term
breaking the $O(4)$ symmetry looks like 
\begin{equation}
\mathcal{L}_{M} = M^2 (\cos\psi - 1).
\end{equation}

The quadratic part of the Lagrangian used for the perturbation expansion can
be selected as follows 
\begin{equation}
\mathcal{L}_{*}^{[2]} = \frac{1}{2}(\partial_{\mu}\omega)^2 + \frac{\omega^2%
}{2} ((\partial_{\mu} \theta)^2 + \sin^2\theta (\partial_{\mu} \varphi)^2) - 
\frac{M^2}{2} \omega^2.  \nonumber
\end{equation}
In terms of the pion fields 
\begin{equation}
\pi^{a} = \omega \times
(\cos\theta,\sin\theta\times(\cos\varphi,\sin\varphi)),
\end{equation}
it takes the standard form 
\begin{equation}
\mathcal{L}_{*}^{[2]} = \frac{1}{2} (\partial_{\mu} \pi^{a})^2 - \frac{M^2}{2%
}(\pi^{a}) ^2,
\end{equation}
whereas the Lagrange measure is simply the Euclidean volume $d^3\pi$. The
difference $\delta \mathcal{L}_{*} = \mathcal{L}_{*} + \mathcal{L}_{M} - \mathcal{L}_{*}^{[2]}$ 
can be considered as a perturbation.

The pion fields can be parameterized in various ways. The problem of ambiguities 
of the transition amplitudes, connected to the arbitrariness of that choice, was 
discussed first in Refs.\cite{CHIK,KAME}. It was shown that on-shell amplitudes do not 
depend on the choice of physical variables. This statement is known as the 
"equivalence theorem". The method proposed by Gasser and Leutwyller \cite{GALE} 
associates the QCD Green functions to amplitudes of the effective
chiral Lagrangian. Using this method, the QCD on- and off-shell amplitudes
can be calculated in a way independent on the parameterization.

The $S$-matrix is invariant with respect to a symmetry group, if both the 
action functional and the Lagrange measure entering the path integral over canonical 
coordinates are invariant. The Liouville measure entering the path integral over 
canonical coordinates and momenta is always "flat", since we work in canonical 
basis. The quantization gives a non-trivial Lagrange measure, however 
(cf. Eq.(\ref{MEAS})). In case of the $O(n)$ non-linear 
sigma model, there is only one parameterization which makes the Lagrange measure 
flat, i.e., $\det || g_{ij} || = 1$. This requirement is useful for development 
of perturbation theory which uses an oscillator basis to convert path 
integrals into the Gaussian form. 
            
The weight factor can always be exponentiated to generate an effective Lagrangian 
$\delta \mathcal{L}_{H}$, in which case $\chi ^{\alpha }=\phi ^{\alpha }$ provides 
desired parameterization also. The exponential parameterization of the pion matrix
\begin{equation}
U({\phi }^{\alpha })= e^{i{\tau }^{\alpha }{\phi }^{\alpha }}  \label{EP}
\end{equation}
gives, in particular, 
\begin{equation}
\delta \mathcal{L}_{H}= - \frac{1}{a^{4}}\ln (\sin ^{2}(\phi )\phi ^{-2})
\end{equation} 
where $a$ is a lattice size. $\delta \mathcal{L}_{H}$ diverges in the continuum limit. 
The non-linear sigma model is not a renormalizable theory, so divergences cannot be 
absorbed into redefinition of $F$ and $M_{\pi }$. Using the mean filed (MF) approximation, it is 
usually possible to keep renormalizations finite. The exponentiation of a variable 
weight factor breaks, in general, selfconsistency of the MF approximation of the 
non-linear sigma model.

Divergences arising from $\delta \mathcal{L}_{H}$ could, however, be compensated by 
divergences coming from higher orders ChPT loops. From this point of view it looks 
naturally to attribute $\delta \mathcal{L}_{H}$ to higher orders ChPT loop expansion 
starting from one loop. The MF approximation for ChPT implies then the tree level 
approximation for the non-linear sigma model with $\delta \mathcal{L}_{H}$ neglected. 
Such an approximation, however, neglects the Haar measure from the start. 
It is therefore hard to expect that such an approximation describes correctly the
high temperature regime where the chiral invariance is supposed to be restored.

The self-consistency of the MF approximation of the non-linear sigma model survives with the
one parameterization only.

{\it The invariance under the chiral transformations admits, within the MF approximation, 
only the parameterization given by Eq.(\ref{NPAR}).}

The parameterization based on the dilatation of $\phi ^{\alpha }$ gives $\delta \mathcal{L}_{H}=0$, 
does not involve the higher orders ChPT loops, and allows to work in the continuum limit with 
finite quantities only.

The effective interaction terms in the Lagrangian which appear due to the presence of 
the Haar measure have been discussed earlier in QCD (see \cite{PARA} and references therein).

The $SU(2)$ group has a finite group volume. The integration range of 
$\omega$ fields is therefore be restricted. According to the current paradigm, 
one can extend the integrals over $\omega$ from $- \infty$ to $+ \infty$ within a perturbation 
theory framework. The modification of the result is connected to the integration 
over large field fluctuations, so the variance has, apparently, a non-perturbative 
nature and does thereby not affect the perturbation series. As a matter of fact, this 
justifies the standard loop expansion in ChPT.

\section{Summary}
\setcounter{equation}{0} 

In this work, we discussed analogy between the second-class constraints systems
and gauge theories with the equivalent structure of gauge generators and 
gauge-fixing conditions. Given the symplectic basis for the constraint 
functions exists globally, the second-class constraints systems can be interpreted 
as gauge invariant systems in the unconstrained phase space. Such systems can be 
quantized using the methods specific for gauge theories.

The second-class constraints $\mathcal{G}_{a}$ split in the symplectic basis
into canonical pairs ($\chi_{A}$,$\Omega_{B}$) satisfying $\{ \chi_{A}, \chi_{B}\} \approx 0$, 
$\{ \Omega_{A}, \Omega_{B}\} \approx 0$, and $\{ \chi_{A}, \Omega_{B}\}
\approx \delta_{AB}$. The constraint functions 
($\chi_{A}$,$\Omega_{B}$) can be transformed further, as discussed in 
Sects. 4 and 5, to fulfill the Poisson bracket relations in the strong sense in an 
entire neighborhood of any give point of the constraint submanifold. The Hamiltonian 
function can also be modified to be identically in involution with the constraints. The new constraints define
the same constraint submanifold, whereas the new Hamiltonian and its first
derivatives coincide with the original ones on the constraint submanifold. The
constrained dynamics is thus not modified.
The new constraint functions $\chi_{A}$ and $\Omega_{A}$ are interpreted as gauge-fixing
conditions and first-class constraints associated to gauge transformations.

We do not provide any criterion for what part of the constraints $\mathcal{G}_{a}$
describes the gauge-fixing conditions and what part describes the
first-class constraints associated to gauge transformations. By contrary,
we argue that transition amplitudes of the quantum theory do not depend on the interpretation 
of $\mathcal{G}_{a}$. 

The Dirac's supplementary conditions ${\hat \Omega}_{A}\Psi = 0$ depend 
on the way the constrains $\mathcal{G}_{a}$ were split. These conditions are 
equivalent to ${\hat \chi}_{A}\Psi^{\prime} = 0$, since the corresponding Wigner 
functions coincide on the constraint submanifold. 
The supplementary conditions for the Wigner functions, furthermore, can be 
made to be explicitly invariant with respect to possible transformations of the constraint 
functions. The ambiguity reflects the freedom in extrapolation of the Wigner 
functions from the constraint submanifold into the unconstrained phase space.

We showed, finally, that the proposed quantization scheme applies to 
an $n-1$-dimensional  spherical pendulum, which represents a mechanical version of
the $O(n)$ non-linear sigma model. For this model, we demonstrated the
existence of an underlying gauge symmetry which is the dilatation of the
coordinates $\phi^{\alpha}$. The constraints 
appearing within the Hamiltonian framework are of the second class. If one 
starts, however, from an equivalent Lagrangian in which the underlying gauge 
symmetry is set up explicitly, the same constraints appear as a gauge-fixing condition and a
constraint associated to the dilatation symmetry. It shows that the
interpretation of second-class constraints is a matter of convention.

For holonomic systems, the quantization method based on construction of the 
gauged model does requires neither auxiliary canonical variables nor extended 
configuration space.

For second-class constraints systems, the underlying gauge symmetries induce 
in the configuration space transformations depending on
velocities and involve auxiliary variables. The holonomic systems admit the natural
gauged counterparts in the original configuration space.

The initial configuration space is, in general, too narrow to reflect the gauge 
invariance of a system described in terms of the generalized Hamiltonian 
dynamics. Gauge invariant quantities involving auxiliary variables, furthermore, 
do not belong to the set of physical observables, as distinct from the usual gauge 
theories. The equivalence of the first-class 
constraints systems with the ordinary gauge systems is therefore physically not 
complete and restricted to systems of point particles under holonomic 
constraints, non-holonomic constraints satisfying the Frobenius' condition, and
systems with one primary constraint only.

The path integral representation for the evolution operator of the $O(n)$ non-linear 
sigma model was constructed in Sect. 9 by solving the constraint equations. 
The equivalence with the quantization methods based on the reduction to the equivalent gauge 
systems was demonstrated.

After finishing this work we got to know about works \cite{NAKA84,NAKA89,NAKA93,NAKA01} 
where projection formalism is discussed. 

\begin{acknowledgements}
M.I.K. and A.A.R. wish to acknowledge kind hospitality at the University of T\"ubingen.
This work is supported by DFG grant No. 436 RUS 113/721/0-2 and
RFBR grant No. 06-02-04004. 
\end{acknowledgements}

\begin{appendix}
\section{Two-constraints form of gauged spherical pendulum}
\setcounter{equation}{0} 

Let us consider another example. The system discussed in Sect. 3 is
equivalent to a system described by a vector $\phi^{i} =
(\phi^{0},\phi^{\alpha})$ on the "light-cone" submanifold $\phi^2 =
\phi^{0}\phi^{0} - \phi^{\alpha}\phi^{\alpha} = 0$. The constraint $%
\phi^{\alpha}\phi^{\alpha} = 1$ is equivalent to the constraint $\phi^{0} =
1 $, while the transformations (\ref{P1}) are equivalent to Lorentz boosts
of the null vector $\phi^{i}$ along the vector $\phi^{\alpha}$, 
\begin{equation}
\phi^{i} \rightarrow \phi^{\prime i} = \Lambda ^{i} _{j}(\theta) \phi^{j} =
\exp(\theta)\phi^{i}  \label{BOOS}
\end{equation}
where the "boost velocity" $v = \tanh(\theta)$. The gauge invariance of the $%
\mathcal{L}_{*}$ with respect to the dilatation reflects gauge invariance of
the system with respect to the Lorentz boosts.

The "light-cone" Lagrangian $\mathcal{L}_{2}$ can be constructed by
considering the requirement of the conditional maximum of the action 
\begin{equation}
\max _{\phi^{\alpha}, {\dot \phi}^{\alpha}} \{ \int{\mathcal{L}_{*}dt}
\}|_{\phi^{\alpha}\phi^{\alpha} = 1} = \max _{\phi^{i}, {\dot \phi}^{i}} \{
\int{\mathcal{L}_{2}dt} \}|_{{\phi}^{2} = 0, {\phi}^{0} = 1}.  \label{GVAR}
\end{equation}
If the $\phi^{\alpha}\phi^{\alpha}$ is treated as a gauge parameter, the
problem simplifies. It is sufficient to require 
\begin{equation}
\max _{\phi^{\alpha}, {\dot \phi}^{\alpha}} \{ \int{\mathcal{L}_{*}dt} \} =
\max _{\phi^{i}, {\dot \phi}^{i}} \{ \int{\mathcal{L}_{2}dt} \}|_{{\phi}^{2}
= 0}.  \label{GV}
\end{equation}

The $\mathcal{L}_{2}$ can be chosen as a straightforward extension of $%
\mathcal{L}_{*}$: 
\begin{equation}
\mathcal{L}_{2} = - \frac{1}{2}G_{ij} (\phi) {\dot \phi}^{i} {\dot \phi}^{j}/%
{((\eta \phi)^2 - \phi ^2)}  \label{LAGC}
\end{equation}
where $\eta = (1,0,...,0)$, 
\begin{equation}
G_{ij}(\phi) = g_{ij} - \frac{(\eta \phi)(\eta_{i} \phi_{j} + \eta_{j}
\phi_{i}) - \phi^2 \eta_{i} \eta_{j} - \phi_{i} \phi_{j}}{(\eta \phi)^2 -
\phi ^2},
\end{equation}
and $g_{ij} = $ diag$(1,-1,...,-1)$, $g^{ij} = g_{ij}$.

The tensor $G_{ij}$ obeys 
\begin{eqnarray}
\eta^{i} G_{ij}(\phi) &=& \phi^{i}G_{ij}(\phi) = 0, \\
G_{ij}(\phi) G^{j}_{k}(\phi) &=& G_{ik}(\phi).
\end{eqnarray}
It is invariant with respect to the dilatation (\ref{BOOS}) and the shifts 
\begin{equation}
\phi^{i} \rightarrow \phi^{\prime i} = \phi^{i} + \epsilon \eta^{i}
\label{SHIF}
\end{equation}
where $\epsilon$ is an arbitrary parameter.

Lagrangian (\ref{LAGC}) is well defined for $\phi^2 \neq 0$. It is invariant
with respect to the dilatation (\ref{LAGC}) and the shifts (\ref{SHIF}).

The $\phi^2$ and the $\eta \phi$ are thus gauge functions. They are not
fixed by equations of motion and can be selected to satisfy admissible
constraints. We consider therefore Lagrangian (\ref{LAGC}) without imposing
any constraints. The initial and final conditions ${\phi}^{2} = 0$ and $\eta
\phi = 1$ are gauge-fixing conditions.

The canonical momenta corresponding to the ${\dot \phi}^{i}$ are defined by 
\begin{equation}
\pi _i = \frac{\partial \mathcal{L}_{c}}{\partial {\dot \phi}^{i}} = -
G_{ij} (\phi) {\dot \phi}^{j}/{((\eta \phi)^2 - \phi^2)}.  \label{CAMA}
\end{equation}
They satisfy the primary constraints 
\begin{equation}
\pi _i - G_{ij} \pi ^{ j} \approx 0  \label{PRIM}
\end{equation}
which are equivalent to two ones: 
\begin{eqnarray}
\Omega_1 &=& \phi \pi \approx 0, \\
\Omega_2 &=& \eta \pi \approx 0.  \label{TPRI}
\end{eqnarray}

The primary Hamiltonian can be obtained with the use of Legendre
transformation: 
\begin{equation}
\mathcal{H} = - \frac{1}{2} {((\eta \phi)^2 - \phi ^2)} G^{ij} (\phi) \pi
_{i}\pi _{j}.
\end{equation}
The Poisson bracket for canonical coordinates and momenta have the form 
\begin{equation}
\{ \phi _{i} , \pi _j \} = g_{ij}.  \label{brac}
\end{equation}
The primary constraints are stable with respect to the time evolution: 
\begin{eqnarray}
\{ \Omega_1, \mathcal{H} \} &=& 0, \\
\{ \Omega_2, \mathcal{H} \} &=& 0.
\end{eqnarray}
The Hamiltonian $\mathcal{H}$ is gauge invariant. The primary constraints
are of the first class: 
\begin{equation}
\{ \Omega_1, \Omega_2 \} = \Omega_2.  \label{XXXX}
\end{equation}
The generators of gauge transformations constitute an algebra.

The relations 
\begin{eqnarray}
\{ \phi^{i} , \Omega_1 \} &=& \phi^{i}, \;\;\;\; \{ \phi^{i} , \Omega_2 \} =
\eta^{i}, \\
\{ \pi _{i} , \Omega_1 \} &=& - \pi_{i}, \; \{ \pi _{i} , \Omega_2 \} = 0
\end{eqnarray}
show that the $\Omega_1$ generates the dilatation of the $\phi^{i}$ and the $%
\pi_{i}$, while the $\Omega_2$ generates time-like shifts of $\phi^{i}$.

The gauge-fixing conditions 
\begin{eqnarray}
\chi_{1} &=& \frac{1}{2}\ln((\eta\phi)^2 - \phi^2), \\
\chi_{2} &=& \eta \phi - 1
\end{eqnarray}
generate the following transformations: 
\begin{eqnarray}
\{ \phi^{i},\chi_{1}\} = 0, \;\;\;\;\;\;\;\;\;\;\;\;\;\;\;\;\;\;\; \{
\phi^{i},\chi_{2}\} = 0,\;\;\;\;&& \\
\{ \pi^{i}, \chi_{1}\} = \frac{\phi^{i} - \eta^{i} \eta \phi}{(\eta \phi)^2
- \phi^2}, \;\;\; \{ \pi^{i},\chi_{2}\} = - \eta^{i}.&&
\end{eqnarray}
They are identically in involution with the Hamiltonian: 
\begin{eqnarray}
\{ \chi_1, \mathcal{H} \} &=& 0, \\
\{ \chi_2, \mathcal{H} \} &=& 0.
\end{eqnarray}

The equations of motion generated by the primary Hamiltonian look like 
\begin{eqnarray}
{\dot \phi^{i}} = \{ \phi^{i} , \mathcal{H} \} = - {((\eta \phi)^2 - \phi^2)}
G^{ij} (\phi) \pi _{j},\;\;\;\;\;&&  \label{DOTQ} \\
{\dot \pi _{i}} = \{ \pi _{i}, \mathcal{H} \} \approx - (\phi _{i} - \eta
_{i} {(\eta \phi)}) {G^{jk} (\phi) \pi _j \pi _k}.&&  \label{DOTP}
\end{eqnarray}

The main inference is that starting from the different Lagrangian, an
equivalent first-class constraints system was constructed. There is no
principal distinction between the system described here and the one
discussed in Sect. 3. In particular, one can solve the constraints $%
\chi_{2} = 0$ and $\Omega_{2} = 0$ to remove the canonically conjugate pair (%
$\phi^{0}$,$\pi_{0}$) from the Hamiltonian. The system of Sect. 3 would be
reproduced then explicitly.

\section{Symplectic basis for first-class constraints}
\setcounter{equation}{0} 

In Sect. 4, an equivalent system of second-class constraints 
satisfying involution relations Eqs.(\ref{TQ1}) in the strong sense in an entire neighborhood
of a given point of the constraint submanifold has been constructed. 
The existence of an equivalent Hamiltonian 
identically in involution with the new constraints has also been demonstrated Eq.(\ref{TQ2}).
The equivalence means that the constraint submanifolds and the phase space flows 
on the constraint submanifolds of the dynamical systems coincide. 

Similar statements for first-class constraints systems are proved in Refs. 
\cite{SCHO49,HENN}. The arguments of Refs.\cite{SCHO49,HENN} cannot be 
extended to second-class constraints without additional assumptions. 
Let us discuss the restrictions.

To replace the constraint functions $\Omega_{A}$ ($A = 1,...,m$) by
equivalent constraint functions ${\tilde \Omega}_{A}$ which are identically
in involution $\{ {\tilde \Omega}_{A}, {\tilde \Omega}_{B} \} = 0$, one can
solve equations $\Omega_{A} = 0$ with respect to first $m$ canonical momenta 
$p_{A} = P_{A}$ where $P_{A}$ are functions of the $n$ canonical coordinates
and the remaining $n - m$ canonical momenta. 
To have the constraints ${\tilde \Omega}_{A} = 0$ resolved, one has to require 
\begin{equation}
\det || \frac{\partial {\tilde \Omega}_{A}}{\partial p_{B}} || \ne 0.
\end{equation}
The phase space has a dimension 
$2n$, $n>m$. The functions ${\tilde \Omega}_{A} = p_{A} - P_{A}$ vanish on
the submanifold $\Omega_{A} = 0$ only. The Poisson bracket $\{ {\tilde \Omega}%
_{A}, {\tilde \Omega}_{B} \}$ vanishes weakly and does not depend on the
first $m$ canonical momenta, therefore it vanishes identically.

The same prescription can be used to construct the functions $\chi_{A}$. 
The Poisson bracket $\{ {\tilde \chi}_{A}, {\tilde \chi}_{B} \}$ vanishes
identically. The new constraint functions satisfy $\det \{ {\tilde \chi}%
_{A}, {\tilde \Omega}_{B} \} \neq 0$.

One can define an equivalent Hamiltonian ${\tilde {\mathcal{H}}}$. Let us
substitute $p_{A} = P_{A}$ to the original Hamiltonian $\mathcal{H}^{\prime}$
of the second-class constraints system. The resulting Hamiltonian ${\tilde {%
\mathcal{H}}}$ is first class with respect to the constraints $\mathcal{G}%
_{a} = 0$, so $\{ {\tilde \Omega}_{A}, {\tilde {\mathcal{H}}} \} \approx 0$.
The difference ${\tilde {\mathcal{H}}} - \mathcal{H}^{\prime}$ vanishes on
the constraint submanifold $\Omega_{A} = 0$. The Hamiltonian ${\tilde {\mathcal{%
H}}}$ does not depend on the first $m$ canonical momenta, so the Poisson
bracket does not depend on these canonical momenta either. The ${\tilde {%
\mathcal{H}}}$ is therefore identically in involution with the ${\tilde
\Omega}_{A}$. The similar procedure can be applied for the ${\tilde \chi}_{A}$. 

We should get
finally constraint functions ${\tilde \chi}_{A}$ and ${\tilde \Omega}%
_{A} $ identically in involution with the ${\tilde {\mathcal{H}}}$.

The above arguments do not apply if some constraint functions do not depend
on $p_{A}$ ($q_{A}$). The bracket $\{ {\tilde \Omega}_{A}, {\tilde \Omega%
}_{B} \}$ can, e.g., be proportional to ${\chi}_{C}$ and ${\chi}_{C}$ can in
turn be independent on $p_{A}$. In such a case, the weak equation $\{ {\tilde \Omega}%
_{A}, {\tilde \Omega}_{B} \} \approx 0$ does not convert into the strong one, 
although the both sides do not depend on $p_{A}$. The similar restrictions appear 
in the construction of ${\tilde \chi}_{A}$ and ${\tilde {\mathcal{H}}}$. The 
systems under holonomic constraints have, in particular, constraints 
$\chi_{A} = 0$ which do not depend on the canonical momenta.

\end{appendix}

\end{document}